\documentstyle[11pt]{article}
       \font\tenmsb=msbm10
       \font\sevenmsb=msbm7
       \font\fivemsb=msbm5
       \catcode`\@=11
       \ifx\amstexloaded@\relax\catcode`\@=\active
       \endinput\else\let\amstexloaded@\relax\fi
       \def\spaces@{\space\space\space\space\space}
       \def\spaces@@{\spaces@\spaces@\spaces@\spaces@\spaces@}
       \def\space@.{\futurelet\space@\relax}
       \space@. %
       \def\Err@#1{\errhelp\defaulthelp@\errmessage{AmS-TeX error: #1}}
       \def\relaxnext@{\let\next\relax}
       \def\accentfam@{7}
       
       \def\noaccents@{\def\accentfam@{0}}
       \def\Cal{\relaxnext@\ifmmode\let\next\Cal@\else
       \def\next{\Err@{Use \string\Cal\space only in math mode}}\fi\next}
       \def\Cal@#1{{\Cal@@{#1}}}
       \def\Cal@@#1{\noaccents@\fam\tw@#1}

       \def\Bbb{\relaxnext@\ifmmode\let\next\Bbb@\else
       \def\next{\Err@{Use \string\Bbb\space only in math mode}}\fi\next}
       
       \def\Bbb@#1{{\Bbb@@{#1}}}
       \def\Bbb@@#1{\noaccents@\fam\msbfam#1}
       \newfam\msbfam
       \textfont\msbfam=\tenmsb
       \scriptfont\msbfam=\sevenmsb
       \scriptscriptfont\msbfam=\fivemsb

\def\N{{\Bbb N}}
\def\Z{{\Bbb Z}}
\def\R{{\Bbb R}}
\def\T{{\Bbb T}}
\def\C{{\Bbb C}}
\def\H{{\Bbb H}}

\newtheorem{Theorem}{Theorem}
\newtheorem{Lemma}{Lemma}[section]
\newtheorem{Proposition}{Proposition}
\newtheorem{Corollary}{Corollary}
\newtheorem{Remark}{Remark}[section]

\newtheorem{Definition}{Definition}[section]

\@addtoreset{equation}{section}

\newcommand{\qed}{\nolinebreak\hfill\rule{2mm}{2mm}
\par\medbreak}
\newcommand{\proof}{\par\medbreak\it Proof: \rm}

\newcommand{\la}{\langle }
\newcommand{\ra}{\rangle }

\newcommand{\beq}{\begin{equation} }
\newcommand{\eeq}{\end{equation} }

\begin{document}
\setlength{\columnsep}{5pt}
\title{Ballistic Motion in One-Dimensional Quasi-Periodic Discrete Schr\"odinger Equation}

\author{Zhiyan Zhao\footnote{This work is supported by ANR grant
``ANR-14-CE34-0002-01'' for the project "Dynamics and CR geometry".}
\\
 {\footnotesize Institut de Math\'{e}matique de Jussieu-Paris Rive Gauche(FSMP contract), 75013 Paris, France}\\
 {\footnotesize Email: zyqiao@gmail.com}}


\date{}
\maketitle

\begin{abstract}
For the solution $q(t)=(q_n(t))_{n\in\Z}$ to one-dimensional discrete Schr\"odinger equation
$${\rm i}\dot{q}_n=-(q_{n+1}+q_{n-1})+ V(\theta+n\omega) q_n, \quad n\in\Z,$$
with $\omega\in\R^d$ Diophantine, and $V$ a real-analytic function on $\T^d$, we consider the growth rate of the diffusion norm $\|q(t)\|_{D}:=\left(\sum_{n}n^2|q_n(t)|^2\right)^{\frac12}$ for any non-zero $q(0)$ with $\|q(0)\|_{D}<\infty$.
We prove that $\|q(t)\|_{D}$ grows {\it linearly} with the time $t$ for any $\theta\in\T^d$ if $V$ is sufficiently small.
\end{abstract}

\tableofcontents

\section{Introduction and main results}\label{sec1}
\noindent

Consider the solution $q(t)$ to one-dimensional discrete Schr\"odinger equation
\begin{equation}\label{target}
{\rm i}\dot{q}_n=-(q_{n+1}+q_{n-1})+V(\theta+ n\omega)q_n, \quad n\in\Z,
\end{equation}
with $V:\T^d\rightarrow \R$ analytic in a complex neighbourhood of $\T^d$ $\{z\in\C^d: |\Im z|<r\leq 1\}$, and $\omega\in\R^d$ Diophantine, i.e., there exist $\gamma>0$, $\tau>d-1$, such that
\begin{equation}\label{dio}
\inf_{j\in\Z}\left|\frac{\la k, \omega\ra}{2}-j\pi\right| > \frac{\gamma}{|k|^{\tau}},\quad \forall k\in\Z^d\setminus\{0\}.
\end{equation}
We want to observe the growth rate with $t$ of the ``diffusion norm"
$$\|q(t)\|_{D}:=\left(\sum_{n\in\Z}n^2 |q_n(t)|^2\right)^\frac12,$$
 provided that $q(0)\neq0$ and $\|q(0)\|_{D}< \infty$.

It is well known that the $\ell^2-$norm $\sum_{n\in\Z}|q_n(t)|^2$ is conserved for Eq.(\ref{target})(see e.g., (2.5) of \cite{BW}).
The initial condition $\|q(0)\|_D<\infty$ indicates the concentration on the lower modes at $t=0$.
The diffusion norm $\|q(t)\|_D$ measures the propagation into higher ones. For more description of the diffusion norm, refer to \cite{BW}.

With the initial condition $\|q(0)\|_D<\infty$, we have $\|q(t)\|_{D}<\infty$ for any finite $t$.
More precisely, we have the general ballistic upper bound(Lieb-Robinson bound\cite{LR})
\begin{equation}\label{ballistic_upper_bound}
\|q(t)\|_{D}\leq \|q(0)\|_{D}+ 2\|q(0)\|_{\ell^2(\Z)} t,
\end{equation}
if the corresponding linear self-adjoint Schr\"odinger operator is bounded.
See also, e.g., Appendix B in \cite{AizenmanW} or Theorem 2.1 in \cite{DamanikLY} for the proof.

Since we are considering the solution of the linear equation (\ref{target}), it is necessary to study the spectral behavior of the linear Schr\"odinger operator $H:\ell^2(\Z)\rightarrow \ell^2(\Z)$,
$$
(Hq)_n=-(q_{n+1}+q_{n-1})+V_n q_n, \quad n\in\Z.
$$
In the case that $H$ has only pure point spectrum, Simon\cite{Simon1990} has shown ``absence of ballistic motion", i.e.,
$$\lim_{t\rightarrow\infty}t^{-1}\|q(t)\|_D=0 \;\ {\rm with} \;\ q(0) \;\ {\rm well-localized},$$
which gives a partial answer to the question of Joel Lebowitz asking if the ballistic motion did not have its roots in absolutely continuous spectrum.
In particular, for the pure point spectrum, the phenomenon ``dynamical localization", which implies boundedness of $\|q(t)\|_D$ for the exponentially decaying initial data, has been well studied and has been proven in many models(refer to \cite{DamanikS, GD, GJ}).

In contrast, the behaviour of solution is totally different in the case that the spectrum of $H$ is purely absolutely continuous.
As shown in RAGE Theorem\cite{CFKS}, it is easy to get the propagation which is related to the growth of $\|q(t)\|_D$.
Corresponding to the question of Joel Lebowitz, the appearance of ``ballistic motion" for Eq (\ref{target}) is quite possible in certain cases of absolutely continuous spectrum.
A time-averaged statement by Guarneri-Combes-Last theorem\cite{Last} shows that, in the presence of absolutely continuous spectrum,
$$
\liminf_{t\to\infty}\frac1T\int_0^T \|q(t)\|_D \, dt\geq C
$$
for some positive constant $C$.
Damanik-Lukic-Yessen\cite{DamanikLY} have recently shown the stronger version of ballistic motion(i.e., the above inequality without time-averaging) for the periodic Schr\"odinger equation, as the periodic Schr\"odinger operator is a well-known example of purely absolutely continuous spectrum. This is
an extension of the work of Asch-Knauf\cite{AschKnauf} for Schr\"{o}dinger operators.

As for the quasi-periodic Schr\"odinger equation, the corresponding linear operator is $H=H_\theta:\ell^2(\Z)\rightarrow \ell^2(\Z)$,
$$
(H_\theta q)_n=-(q_{n+1}+q_{n-1})+V(\theta+n\omega) q_n, \quad n\in\Z,
$$
with $V$ and $\omega$ given as in (\ref{target}).
It is well known that the spectrum of $H_\theta$, which
we shall denote by $\sigma(H_\theta)$ or simply $\sigma(H)$, is a closed non-empty subset of the interval $[-2-|V|_r,\, 2+|V|_r]$.
It will be shown that the spectrum is purely absolutely continuous when $V$ is small enough(see Proposition \ref{spec_ac} in Subsection \ref{ac_spec}).
For this model, Kachkovskiy\cite{Kachkovskiy} has proven a time-averaged version of ballistic transport for a subsequence of times, provided that $V$ is small enough. In particular, the same conclusion is shown if $H$ has purely absolutely continuous spectrum with one Diophantine frequency.

In this paper, for the quasi-periodic Schr\"{o}dinger equation (\ref{target}), a rigorous proof for the linear growth of the diffusion norm will be given, corresponding to a numerical result \cite{HiAbe} for Harper's model.

\begin{Theorem}\label{thm-qp}
Consider the solution $q(t)$ to Eq.(\ref{target}).
There exists an $\varepsilon_*=\varepsilon_*(\gamma,\tau,r)$, such that
if $|V|_r=\varepsilon_0<\varepsilon_*$, then for any $\theta\in\T^d$,
there is a constant $0< C< 3\|q(0)\|_{\ell^2(\Z)}$, depending on $\varepsilon_0$, $\theta$ and $q(0)$,
such that, for some numerical constant $0<\zeta<1$,
$$
 \liminf_{t\rightarrow\infty}t^{-1}\|q(t)\|_{D} \geq \frac{C}{1+\varepsilon_0^\zeta}, \quad \limsup_{t\rightarrow\infty}t^{-1}\|q(t)\|_{D} \leq \frac{C}{1-\varepsilon_0^\zeta}.
$$
\end{Theorem}

\noindent {\bf Idea of proof.} The main strategy is to relate the linear growth of diffusion norm to the spectral transformation of the solution $q(t)$.
Roughly speaking, for $g(E,t)=\sum_{n}q_n(t)\psi_n(E)$, with $(\psi_n(E))_n$, $E\in\sigma(H)$, a generalized eigenvector of $H$, we have that it satisfies ${\rm i}\partial_t g(E,t) = E g(E,t)$, then
$$\sum_{n}q_n(t)\psi_n(E)=g(E,t) = e^{-{\rm i}E t}g(E,0).$$
So if $\psi_n(E)$ has nice differentiability and the derivative is well estimated, we can get
$$\sum_{n}q_n(t)\psi'_n(E)=\partial_E g(E,t)\sim t.$$
If, with some suitable measure $d\varphi$ supported on $\sigma(H)$, we have
$$\left\|\sum_{n}q_n(t)\psi'_n\right\|_{L^2(d\varphi)}\sim\left( \sum_{n\in\Z}n^2|q_n(t)|^2\right)^{\frac12},$$
the linear growth of $\|q(t)\|_D$ is shown.

The above process is realized by the ``modified spectral transformation", which is
 written with the same formulation as that of Coddington-Levinson\cite{CL} for the classical spectral transformation.
The generalized eigenvectors, with the Bloch-wave structures, are constructed by the previous works of Eliasson\cite{E92} and Hadj Amor\cite{HA} for the reducibility of Schr\"odinger cocycle.
By adding some smoothing factors to the generalized eigenvectors(in a small part of the spectrum), the differentiability is improved.
Moreover, the classical spectral measure, which was introduced by the $m-$functions, is replaced by some suitable measure according to the transversality of the rotation number of Schr\"odinger cocycle.
In this way, the $L^2-$norm of the derivative (w.r.t. $E$) of the modified spectral transformation is close to the diffusion norm.

%

\section{Preliminaries and notations}\label{sec_pre_notation}

\subsection{Schr\"{o}dinger operator and Schr\"{o}dinger cocycle}\label{pre_op_cocycle}
\noindent

 In this subsection, we recall some basic notions and well-known results for the quasi-periodic Schr\"{o}dinger operator
  $H=H_\theta:\ell^2(\Z)\rightarrow \ell^2(\Z)$,
$$
(H q)_n=-(q_{n+1}+q_{n-1})+V(\theta+n\omega) q_n, \quad n\in\Z,
$$
with $V$ and $\omega$ given as in (\ref{target}), and the corresponding Schr\"{o}dinger cocycle $(\omega, A_0+F_0)$:
\begin{equation}\label{qpcocycle}
\left(\begin{array}{c}
q_{n+1} \\[1mm]
q_{n}
\end{array}
\right)=(A_0(E)+F_0(\theta+n\omega))\left(\begin{array}{c}
q_{n} \\[1mm]
q_{n-1}
\end{array}
\right),
\end{equation}
with $A_0(E):=\left(\begin{array}{cc}
            -E & -1 \\
            1 & 0
          \end{array}
\right)$ and $F_0(\theta):=\left(\begin{array}{cc}
            V(\theta) & 0 \\
            0 & 0
          \end{array}
\right)$. Note that $(\omega, A_0+F_0)$ is equivalent to the eigenvalue problem $Hq=Eq$.

\subsubsection{Spectral measure and integrated density of states}
\noindent

Fixing any phase $\theta\in\T^d$ and any $\psi\in\ell^2(\Z)$, let $\mu_\theta=\mu_{\theta,\psi}$ be
the spectral measure of $H = H_\theta$ corresponding to $\psi$, which is defined so that
$$\la(H_\theta-E)^{-1}\psi, \psi \ra = \int_\R\frac{1}{E-E'}d\mu_{\theta,\psi}(E'), \quad \forall \, E\in \C\setminus\sigma(H).$$
From now on, we restrict our consideration to $\mu_\theta=\mu_{\theta, e_{-1}}+\mu_{\theta, e_0}$ and just call it the {\bf spectral measure}, where
$\{e_n\}_{n\in\Z}$ is the canonical basis of $\ell^2(\Z)$.
Since $\{e_{-1}, e_0\}$ forms a generating basis of $\ell^2(\Z)$\cite{CarmonaLacroix},
that is, there is no
proper subset of $\ell^2(\Z)$ which is invariant by $H$ and contains $\{e_{-1}, e_0\}$. In particular
the support of $\mu_\theta$ is $\sigma(H)$ and if $\mu_{\theta}$ is absolutely continuous then any $\mu_{\theta,\psi}$ , $\psi\in\ell^2(\Z)$, is absolutely continuous.

The {\bf integrated density of states} is the function $k:\R\rightarrow[0,1]$ such that
$$k(E)=\int_{\T^d} \mu_\theta(-\infty,E] \,  d\theta,$$
which is a continuous non-decreasing surjective function.

\subsubsection{Rotation number and Lyapunov exponent}
\noindent

Related to the Schr\"odinger cocycle $(\omega, A_0+F_0)$, a unique representation can be given for the rotation number
$\rho=\rho_{(\omega, A_0+F_0)}$.
Indeed, the rotation number is defined for more general quasi-periodic cocycles.
It is introduced originally by Herman\cite{Herman} in this
discrete case(see also Delyon-Souillard\cite{DelyonSouillard}, Johnson-Moser\cite{JM}, Krikorian\cite{Kri}).
For the precise definition, we follow the same presentation as in \cite{HA}.

Given $\displaystyle A: \T^d\mapsto SL(2,\R)$ continuous with
$A(\theta)=\left(\begin{array}{cc}
a(\theta) & b(\theta) \\
c(\theta) & d(\theta)
\end{array}\right)$,
we define the map
$$
\begin{array}{llll}
T_{(\omega,\, A)}:& \displaystyle  \T^d\times \frac12\T &\rightarrow& \displaystyle  \T^d\times \frac12\T \\[3mm]
   &(\theta,\varphi)&\mapsto&(\theta+\omega, \, \phi_{(\omega,\, A)}(\theta,\varphi))
\end{array},$$
where $\frac12\T:=\R/\pi\Z$ and $\phi_{(\omega,\, A)}(\theta,\varphi)=\arctan\left(\frac{c(\theta)+d(\theta)\tan\varphi}{a(\theta)+b(\theta)\tan\varphi}\right).$
Assume that $A(\theta)$ is homotopic to the identity, then the same is true for the map $T_{(\omega,\, A)}$ and therefore it admits a continuous lift
$$
\begin{array}{llll}
\tilde T_{(\omega,\, A)}:& \displaystyle  \T^d\times \R &\rightarrow& \displaystyle  \T^d\times \R \\[2mm]
   &(\theta,\varphi)&\mapsto&(\theta+\omega, \, \tilde\phi_{(\omega,\, A)}(\theta,\varphi))
\end{array}$$
such that $\tilde\phi_{(\omega,\, A)}(\theta,\varphi)\; {\rm mod}\; \pi=\phi_{(\omega,\, A)}(\theta,\varphi\; {\rm mod}\; \pi)$.  The function $$(\theta,\varphi) \mapsto \tilde\phi_{(\omega,\, A)}(\theta,\varphi)-\varphi$$ is $(2\pi)^d-$periodic in $\theta$ and $\pi-$periodic in $\varphi$.
We define now $\rho(\tilde{\phi}_{(\omega,\, A)})$ by
$$\rho(\tilde{\phi}_{(\omega,\, A)})=\limsup_{n\rightarrow+\infty}\frac1n (p_2\circ \tilde T^n_{(\omega,\, A)}(\theta,\varphi)-\varphi )\in\R,$$
where $p_2(\theta,\varphi)=\varphi$.
This limit exists for all $\theta\in\T^d$, $\varphi\in\R$, and the convergence is uniform in $(\theta,\varphi)$(For the existence of this limit and its properties we can refer to \cite{Herman}).
The class of number $\rho(\tilde\phi_{(\omega,\, A)})$ in $\frac12\T$, which is independent of the chosen lift, is called the {\bf rotation number} of the skew-product system
$$
\begin{array}{llll}
(\omega, A):& \displaystyle  \T^d\times \R^2 &\rightarrow& \displaystyle  \T^d\times \R^2 \\[2mm]
   &(\theta,\, y)&\mapsto&(\theta+\omega,\, A(\theta)y)
\end{array},$$
and we denote it by $\rho_{(\omega,\, A)}$.
For more elementary properties, refer to Appendix of \cite{HA}.

For the quasi-periodic cocycle
$\left(\begin{array}{c}
q_{n+1} \\[1mm]
q_{n}
\end{array}
\right)=A(\theta+n\omega)\left(\begin{array}{c}
q_{n} \\[1mm]
q_{n-1}
\end{array}
\right)$,
with $\displaystyle A: \T^d\mapsto SL(2,\R)$ continuous and $\omega=(\omega_1,\cdots,\omega_d)\in\R^d$ rationally independent,
the {\bf Lyapunov exponent} $L=L_{(\omega,\, A)}$ is defined by
$$L_{(\omega,\, A)}:=\lim_{n\rightarrow \infty}\frac{1}{n}\int_{\T^d}\ln|A(\theta+n\omega)\cdots A(\theta+\omega)|\, d\theta.$$
By Kingman's subadditive ergodic theorem,
$$L_{(\omega,\, A)}:=\lim_{n\rightarrow \infty}\frac{1}{n}\ln|A(\theta+n\omega)\cdots A(\theta+\omega)|.$$

In particular, for quasi-periodic Schr\"{o}dinger cocycle $(\omega, A_0+F_0)$ given in (\ref{qpcocycle}), a well-known result of Kotani theory shows, if the linear Schr\"{o}dinger operator $H$ has purely absolutely continuous spectrum, then $L(E)=0$ a.e. on $\sigma(H)$.
Moreover,
the Thouless formula relates the Lyapunov exponent to the integrated density of states:
$$L(E)=L_{(\omega, A_0+F_0)}(E) = \int_{\R} \ln|E'-E| \, dk(E').$$
There is also a relation between the rotation number and  the integrated density of states:
$$k(E)=\left\{ \begin{array}{cl}
                 0, & E\leq \inf\sigma(H) \\[1mm]
                \frac{\rho(E)}{\pi}, & \inf\sigma(H)<E<\sup\sigma(H) \\[1mm]
                 1, & E \geq \sup\sigma(H)
               \end{array}
 \right.  .$$
By the gap-labelling theorem(see, e.g., \cite{DelyonSouillard, JM}), $k(E)=\frac{\rho(E)}{\pi}$ is constant in a gap of $\sigma(H)$(i.e., an interval on $\R$ in the resolvent set of $H$), and each gap is labelled with $l\in\Z^d$ such that $\rho=\frac{\la l,\, \omega\ra}{2}$ mod $\pi$ in this gap.

\subsubsection{The $m-$functions}
\noindent

The spectral measure $\mu=\mu_\theta$ can be studied through
its Borel transform $M = M_\theta$:
$$M(z)=\int\frac{1}{E'-z}d\mu(E').$$
It maps the upper-half plane $\H:=\{z\in\C: \Im z>0\}$ into itself.

From the limit-point theory, for $z\in\H$, there are two solutions $u^{\pm}$, with $u_0^{\pm}\neq 0$, which are $\ell^2$ at $\pm\infty$ and satisfying $Hu^{\pm} = zu^{\pm}$,
defined up to normalization.
Let
$m^{\pm}:=-\frac{u^{\pm}_{\pm1}}{u^{\pm}_{0}}$.
$m^+$ and $m^-$ are Herglotz functions, i.e., they map $\H$ holomorphically into itself(see, e.g., \cite{Simon1983} for more properties of Herglotz function).
Moreover, it is well known that
$$M =\frac{m^+m^- -1}{m^+ + m^-}.$$
By the property of Herglotz function, we know that for almost every $E\in\R$,
the non-tangential limits $\lim_{\epsilon\rightarrow0}m^{\pm}(E+{\rm i}\epsilon)$ exist, and they
define measurable functions on $\R$ which we still denote $m^\pm(E)$.

We have the following key result of Kotani Theory\cite{Simon1983}.
\begin{Lemma}[Theorem 2.2 of \cite{Avila}]
For every $\theta$, for a.e. $E$ such that $L(E) = 0$, we have
$m^+(E)=m^-(E)$.
\end{Lemma}

\subsubsection{Classical spectral transformation}
\noindent

%

Let $u(E)$ and $v(E)$ be the solutions of the eigenvalue problem $Hq=Eq$ such that
$\left(
\begin{array}{cc}
u_1 & v_1 \\
u_0 & v_0 \\
\end{array}
\right)=\left(
\begin{array}{cc}
1 & 0 \\
0 & 1 \\
\end{array}
\right)
$.
We have
\begin{Theorem}[Chapter 9 of \cite{CL}]\label{spectral_measure_matrix}
There exists a non-decreasing Hermitian matrix $\mu=(\mu_{jk})_{j,k=1,2}$ whose elements are of bounded variation on every finite interval on $\R$, satisfying
$$\mu_{jk}(E_2)-\mu_{jk}(E_1)=\lim_{\epsilon\rightarrow 0_+}\frac{1}{\pi}\int_{E_1}^{E_2}\Im M_{jk}(\nu+i\epsilon)d\nu,$$
at points of continuity $E_1$, $E_2$ of $\mu_{jk}$, where on $\H$,
$$M=\left(\begin{array}{cc}
            M_{11} & M_{12} \\
            M_{21} & M_{22}
          \end{array}
\right):=-\frac{1}{m^{+}+m^{-}}\left(\begin{array}{cc}
                                  1 & m^+ \\
                                  -m^- & -m^+m^-
                                \end{array}
\right),$$
such that for any $q\in \ell^2(\Z)$, with
$(g_1(E),g_2(E)):=\left(\sum_{n\in\Z} q_n u_n(E), \sum_{n\in\Z} q_n v_n(E)\right)$,
we have Parseval's equality
$$\sum_{n\in\Z}|q_n|^2=\int_{\R}\sum_{j,k=1}^2 \bar g_j(E) g_k(E) d\mu_{jk}(E).$$
\end{Theorem}

Given any matrix of measures on $\R$ $d\varphi=\left(\begin{array}{cc}
                 d\varphi_{11} & d\varphi_{12} \\[1mm]
                 d\varphi_{21} & d\varphi_{22}
               \end{array}\right)$,
let ${\cal L}^2(d\varphi)$ be the space of vectors
$G=(g_j)_{j=1,2}$, with $g_j$ functions of $E\in\R$ satisfying
\begin{equation}\label{gen_L2}
\|G\|_{{\cal L}^2(d\varphi)}^2:=\sum_{j,k=1}^2 \int_\R g_j \, \bar g_k \,  d\varphi_{jk}<\infty.
\end{equation}
In view of Theorem \ref{spectral_measure_matrix}, the map
$(q_n)_{n\in\Z}\mapsto \left(\begin{array}{c}
                                                         \sum_{n\in\Z}q_n u_n(E) \\[1mm]
                                                         \sum_{n\in\Z}q_n v_n(E)
                                                       \end{array}
     \right)$
defines a unitary transformation between $\ell^2(\Z)$ and ${\cal L}^2(d\mu)$.
We call it as the {\bf classical spectral transformation}.

By Chapter $\uppercase\expandafter{\romannumeral5}$ of \cite{PF}(Page 297), we know that
the matrix of measures $(d\mu_{jk})_{j,k=1,2}$ is Hermitian-positive, and
therefore each $d\mu_{jk}$ is absolutely continuous with respect to the measure $d\mu_{11}+d\mu_{22}$.
This measure is absolutely continuous with respect to the above spectral measure $\mu_\theta=\mu_{\theta, e_{-1}}+\mu_{\theta, e_0}$
and it determines the spectral type of the operator.
In particular, if the spectrum of $H$ is purely absolutely continuous, we have, for any $q\in\ell^2(\Z)\setminus\{0\}$,
the classical spectral transformation is supported on a subset of $\sigma(H)$ with positive Lebesgue measure.

\

For the classical spectral transformation, there are some singularities with respect to $E$. More precisely, $u_n$ and $v_n$ are not well differentiated somewhere in the spectrum $\sigma(H)$.
For example, for the free Schr\"{o}dinger operator
$(H q)_n=-(q_{n+1}+q_{n-1})$,
we have $\sigma(H)=[-2,2]$ and for $E\in\sigma(H)$ the rotation number is $$\xi_0(E)=\rho_{(\omega, A_0)}(E)=\cos^{-1}\left(-\frac{E}{2}\right)\in[0,\pi].$$
Since $-E=2\cos\xi_0$, we can see that the two generalized eigenvectors
\begin{equation}\label{classical_uv}
u_n=\frac{\sin n\xi_0}{\sin \xi_0}, \quad v_n=-\frac{\sin (n-1)\xi_0}{\sin \xi_0}
\end{equation}
satisfy $\left(
\begin{array}{cc}
u_1 & v_1 \\
u_0 & v_0 \\
\end{array}
\right)=\left(
\begin{array}{cc}
1 & 0 \\
0 & 1 \\
\end{array}
\right)
$
and, on $(-2,2)$, $\xi_0'=\frac{1}{2\sin\xi_0}$.
Differentiating $u_n$, we have
$$u'_n=\frac{1}{2\sin\xi_0}\left( \frac{n\cos n\xi_0}{\sin \xi_0}-\frac{\sin n\xi_0\cdot\cos\xi_0}{\sin^2 \xi_0} \right).$$
The singularity comes when $\xi_0$ approaches $0$ and $\pi$.

\subsection{Regularity in the sense of Whitney}
\noindent

Given a closed subset $S$ of $\R$. We give a precise definition of ${\cal C}^1$ in the sense of Whitney, corresponding to a more general definition in \cite{Poschel}.

\begin{Definition}\label{definition_whitney}
Given two functions $F_0$, $F_1:S\rightarrow \C$(or $SL(2,\C)$) with some $0<M<\infty$, such that
\begin{equation}\label{condition_Whitney}
|F_0(x)|,\, |F_1(x)|\leq M, \;\  |F_0(x)-F_0(y)-F_1(y)(x-y)|\leq M|x-y|, \quad  \forall \, x,y\in S.
\end{equation}
We say that $F_0$ is {\bf ${\cal C}^1$ in the sense of Whitney} on $S$, denoted by $F_0\in{\cal C}_W^1(S)$, with the first order derivative $F_1$.
The ${\cal C}_W^1(S)-$norm of $F_0$ is defined as
$$|F_0|_{{\cal C}_W^1(S)}:=\inf M.$$
\end{Definition}

\begin{Remark}
By Whitney's extension theorem\cite{Whitney}, we can find an extension $\tilde F:\R\rightarrow \C$, which is ${\cal C}^1$ on $\R$ in the natural sense, such that
$\tilde F|_S=F_0$ and $\tilde F'|_S=F_1$.
\end{Remark}


\subsection{Notations}

\noindent {\rm 1)} With $\omega$ the Diophantine vector as above, we denote $\la k\ra:=\frac{\la k,\,  \omega\ra}{2}$ mod $\pi$ for any $k\in\Z^d$, and $|\cdot|$ over $\la k\ra$, $\rho-\la k\ra$, etc. is always modulated into $[0, \frac{\pi}{2}]$ as in (\ref{dio}).

\smallskip

\noindent {\rm 2)} For any subset $S\subset\R$, let $\sharp(S)$ denote its cardinality of set, $\partial S$ be the set of its endpoints, $|S|$ be its Lebesgue measure, $\overline S$ be its closure, and $\rho(S)$ be its image by $\rho=\rho_{(\omega,\, A_0+F_0)}$.
 \begin{itemize}
   \item Given any function $F$ on $S \times(2\T)^d$, possibly matrix-valued, let $$|F|_{S,\, (2\T)^d}:=\sup_{E\in S}\sup_{\theta\in(2\T)^d}|F(E,\theta)|.$$
If $F$ is ${\cal C}_W^1$ on $S$, then we define
    $|F|_{{\cal C}_W^1(S),\, (2\T)^d}:=\sup_{\theta\in(2\T)^d}|F(\cdot,\theta)|_{{\cal C}_W^1(S)}$.
   \item If $F$ is left and right continuous on $E$, then
   $F(E\pm):=\lim_{\epsilon\rightarrow0+}F(E\pm\epsilon)$.
   On the interval $(E_1, E_2)\subset\R$, if $F$ is left and right continuous on $E_1$ and $E_2$, then
   $$\left. F\right|_{[E_1, E_2]}=\left. F\right|_{(E_1, E_2)}:=F(E_{2}-)-F(E_1+),\; \left. F\right|^{E^+_{2}}_{E^-_1}:=F(E_{2}+)-F(E_1-).$$
 \end{itemize}

\smallskip

\noindent {\rm 3)} For the quantities depending on $E\in\R$, we do not always present this dependence explicitly and we simplify the notation ``$\partial_E$" into $\partial$, which denotes the derivative in the sense of Whitney on a certain subset of $\R$.

\smallskip

\noindent {\rm 4)} For any $n\in\Z$, $n_\Delta$ varies among $n$ and $n\pm1$, and $\delta_{n,n_\Delta}:=\left\{\begin{array}{cc}
1,&n_\Delta=n\\[1mm]
0,&n_\Delta\neq n
\end{array}\right.$.

\section{Reducibility of Schr\"odinger cocycle and its applications}\label{spectral_properties}
\noindent

Based on the general notions for Schr\"odinger operator and Schr\"odinger cocycle given in the previous section,
we present some further spectral properties, under the assumption that the potential function $V$ is sufficiently small.

\subsection{KAM scheme for the reducibility}
\noindent

In this subsection, we review the KAM theory of Eliasson\cite{E92} and Hadj Amor\cite{HA} for the reducibility of Schr\"odinger cocycle.
This work relates the reducibility and the rotation number $\rho=\rho_{(\omega, \, A_0+F_0)}$
globally, and it improves the previous works of Dinaburg-Sinai\cite{DiSi} and Moser-P\"oschel\cite{MP}.

With $\varepsilon_0=|V|_r$, $\sigma=\frac{1}{200}$, define the sequences as in \cite{HA}:
$$\varepsilon_{j+1}=\varepsilon_{j}^{1+\sigma}, \; \; N_j=4^{j+1}\sigma|\ln\varepsilon_j|, \quad  j\geq 0.$$

\begin{Proposition} \label{propsana} There exists $\varepsilon_*=\varepsilon_*(\gamma,\tau, r)$ such that if $|V|_r=\varepsilon_0\leq\varepsilon_*$,
 then there is a full-measure subset $\Sigma=\cup_{j\geq0}\Sigma_j$ of $\sigma(H)$ with $\{\Sigma_{j}\}_j$ mutually disjoint Borel sets, satisfying
 $$|\rho\left(\Sigma_{j+1}\right)|\leq |\ln\varepsilon_0|^{(j+1)^3 d}\,  \varepsilon_{j}^{\sigma}, \quad  j\geq 0,$$
such that the following statements hold.
 \begin{itemize}
\item [(1)] The Schr\"odinger cocycle $(\omega, A_0+F_0)$ is {\bf reducible} on $\Sigma$. More precisely,
there exist
$\left\{ \begin{array}{l}
            B:\Sigma\rightarrow SL(2,\R) \;\  with \;\ eigenvalues \;\  e^{\pm{\rm i}\rho}\\[1mm]
            Z:\Sigma\times(2\T)^d\rightarrow SL(2,\R) \;\ analytic \;\  on \;\ (2\T)^d
          \end{array}
 \right.$
such that
$$Z(\theta+\omega)^{-1} (A_0+F_0(\theta)) \, Z(\theta)=B \ on \ \Sigma.$$
\item [(2)] For every $j\geq 0$, there is $k_j:\Sigma\rightarrow\Z^d$, such that
\begin{itemize}
  \item $|k_l|_{\Sigma_j}=0$ if $l\geq j$,
  \item $0<|k_j|\leq N_{j}$ on $\Sigma_{j+1}$ and $0<|\rho-\sum_{l\geq0}\la k_l\ra|_{\Sigma_{j+1}}< 2 \varepsilon_{j}^{\sigma}$.
\end{itemize}
\item [(3)] $B$ and $Z$ are ${\cal C}^1_W$ on ${\Sigma}_0$, and, with $\xi:=\rho-\sum_{j\geq0}\la k_j\ra$, $s\geq 2$, $\sin^{s+2}\xi\cdot B$ and $\sin^{s+2}\xi\cdot Z$ are ${\cal C}^1_W$
 on each ${\Sigma}_{j+1}$, $j\geq 0$. Moreover,
\begin{equation}\label{limit_state_whitney}
\left\{\begin{array}{lll}
|Z-Id.|_{{\cal C}^1_W({\Sigma}_0),\, (2\T)^d}, & |B-A_0|_{{\cal C}^1_W({\Sigma}_0)}\leq \varepsilon_0^{\frac13} & \\[1mm]
|\sin^{s+2\nu}\xi\cdot Z|_{{\cal C}^\nu_W({\Sigma}_{j+1}),\, (2\T)^d}, & |\sin^{s+2\nu}\xi\cdot B|_{{\cal C}^\nu_W({\Sigma}_{j+1})}\leq \varepsilon_j^{\frac{2\sigma}{3}}, & \nu=0,1
\end{array}\right. .
\end{equation}
 \end{itemize}
\end{Proposition}

\begin{Remark}
The conclusion of Eliasson is originally stated as: the cocycle $(\omega,\, A_0+F_0)$ is reducible if the rotation number $\rho$ is Diophantine
or rational with respect to $\frac{\omega}{2}$.
Here, ``rational w.r.t. $\frac{\omega}{2}$" means $\rho=\la k \ra$ for some $k\in\Z^d$. By the gap-labelling theorem, this case corresponds to the energies in $\R\setminus\sigma(H)$, where the uniform hyperbolicity implies the reducibility.
In contrast, ``Diophantine w.r.t. $\frac{\omega}{2}$" means there exist $\gamma$, $\tau>0$ such that $|\rho-\la l\ra|>\frac{\gamma}{|l|^\tau}$ for any $l\in\Z^d\setminus\{0\}$. This corresponds to the energies in a full-measure subset of $\sigma(H)$.
\end{Remark}

\begin{Remark}\label{rmq_Prop1}
We can call $\Sigma_j$ the {\bf $j^{\rm th}-$level resonance set}.
Associated with the above Diophantine condition, if, in $\sigma(H)$, the rotation number $\rho$ is well separated from $\{\la l\ra\}_{l\in\Z^d\setminus\{0\}}$, it is the idealest case for applying the KAM scheme.
\begin{itemize}
  \item On $\Sigma_0$, there is no resonance for the rotation number $\rho$, so the standard KAM iteration is always applicable.
 $\Sigma_0$ is exactly the positive-measure subset of parameters for reducibility in the result of Dinaburg-Sinai\cite{DiSi}.
  \item On $\Sigma_{j+1}$, $j\geq 0$, there is always a vector $k\in\Z^d$ with $0<|k|\leq N_{j+1}$, which appears as $k=\sum_{l=0}^j k_l$, such that
$0<|\rho-\la k\ra|_{\Sigma_{j+1}}< 2 \varepsilon_{j}^{\sigma}$.
But the resonance stops exactly at the $j^{\rm th}-$KAM step.
We could also apply the standard KAM on these subsets from the $(j+1)^{\rm th}-$step, because we could renormalize $\rho$ into $\xi:=\rho-\la k\ra$(the renormalization is done step by step), which is well separated from $\{\la l\ra\}_{l\in\Z^d\setminus\{0\}}$.
Note that the ``renormalized rotation number" $\xi$ is close to $0$ on $\Sigma_{j+1}$ and it vanishes on the gap of spectrum where $\rho=\la k\ra$. So it can serve as a ``smoothing factor" on $\Sigma_{j+1}$.
\end{itemize}
Because of the difference between the procedures on $\Sigma_0$ and $\Sigma_{j+1}$, the transformation $Z$ and the reduced matrix $B$ possess different properties. In particular, on $\Sigma_{j+1}$, there are singularities like $\sim\sin^{-1}\xi$(and $\sim\sin^{-3}\xi$ after the derivation) for $Z$ and $B$.
Then, by multiplying $\sin^{s}\xi$, $s\geq4$ the regularity is well improved as in (\ref{limit_state_whitney}).
Indeed, to get the ${\cal C}^1_W$ regularity, $\sin^3\xi$ is enough, and the $4^{\rm th}$ power makes the norms small.
For better regularity, higher power of $\sin\xi$ is needed.
\end{Remark}

\begin{Remark}
It has been shown in \cite{E92} and \cite{HA} that, for any $E\in\sigma(H)$, the Schr\"odinger cocycle $(\omega, A_0+F_0)$ is {\bf almost reducible}, i.e.,
we can transform it arbitrarily close to a constant cocycle by a sequence of conjugations, without verifying the convergence of this sequence. On $\Sigma_j$, $j\geq 0$, since the resonance stops at exactly the $j^{\rm th}-$step and afterwards the conjugations are all close to identity, the convergence of sequence of conjugations is shown. Hence, in particular, reducibility holds for a.e. $E\in\sigma(H)$.
\end{Remark}

From now on, we always assume that $|V|_r=\varepsilon_0$ is small enough such that it is compatible with every simple calculation in this paper, e.g., $\varepsilon_0^{\sigma}<\frac{\gamma}{2}$ in (\ref{Diophantine_new}).

\smallskip

We present the proof of the arguments (1) and (2) here, and leave the proof of (3) in Appendix \ref{proofP13}.

\noindent{\bf Proof of Proposition \ref{propsana}(1) and (2):} The proof is written in the following two parts. Some details, which is useful for proving (3) but maybe not directly related to (1) and (2), are also given in this proof.

\smallskip

\begin{itemize}
\item  [Part 1.] KAM scheme for the reducibility of Schr\"odinger cocycle $(\omega,\, A_0+F_0)$
\end{itemize}

\begin{itemize}
  \item [1)] The first step
\end{itemize}

At the initial state $\tilde A_0+\tilde F_0:= A_0+F_0$, we have
$$\tilde A_0=\left(\begin{array}{cc}
-E & -1 \\
1 & 0
\end{array}\right)=C_{\tilde A_0}\left(\begin{array}{cc}
e^{{\rm i}\tilde \alpha_0} & 0 \\
0 & e^{-{\rm i}\tilde \alpha_0}
\end{array}\right)C_{\tilde A_0}^{-1},\quad |\partial^\nu \tilde F_0|_{\T^d}\leq\varepsilon_0, \;\ \nu=0,1,2, $$
where $C_{\tilde A_0}$ is the matrix of normalized eigenvectors of $\tilde A_0$.
The constant cocycle $(\omega,\tilde A_0)$ corresponds to the free Schr\"{o}dinger eigenvalue problem $-(q_{n+1}+q_{n-1})=Eq_n$, and its rotation number on $[\inf\sigma(H), \sup\sigma]$ is given by
\begin{equation}\label{xi_0}
\xi_0=\Re\tilde \alpha_0=\left\{\begin{array}{cl}
                 0, & \inf\sigma(H)\leq E< -2 \\[1mm]
                 \cos^{-1}\left(-\frac{E}{2}\right), & -2\leq E \leq 2 \\[1mm]
                 \pi, & 2<E\leq \sup\sigma(H)
               \end{array}
 \right. .
\end{equation}
 $\xi_0$ is non-decreasing on $\R$ and
 \begin{equation}\label{partial_xi_0}
 \partial\xi_0=\frac{1}{\sqrt{4-E^2}}=\frac{1}{2\sin\xi_0} \ {\rm on} \  (-2,2).
 \end{equation}
 So $E=\pm 2$ are the only two singularities of $\xi_0$.
 It is direct to see that $\partial\xi_0>\frac13$ on $(-2,2)$.

The first step is to transform $\tilde A_0+ \tilde F_0(\theta)$ into $\tilde A_{1}+\tilde F_{1}(\theta)$ with $|\partial^\nu \tilde F_1|_{\T^d}\leq\varepsilon_1$, $\nu=0,1,2$, and the property of $\tilde A_{1}$ similar to that of $\tilde A_{0}$.
As shown in Proposition 2 of \cite{HA}, to carry out the standard KAM step, we need the small divisor condition
\begin{equation}\label{Diophantine_condition_0}
|\xi_0 -\la k \ra|\geq \frac{\varepsilon_0^{\sigma}}{|k|^\tau}, \quad \forall \,  0<|k|\leq N_0.
\end{equation}
Related to this condition, there are two cases about the construction of the transformation.
\begin{itemize}
  \item  Case 1. For some $0<|k_0|\leq N_0$,  the condition (\ref{Diophantine_condition_0}) does not hold, i.e.,
  \begin{equation}\label{resonance_condition_0}
  |\xi_0 -\la k_0 \ra|< \frac{c\varepsilon_0^{\sigma}}{|k_0|^\tau},
  \end{equation}
  with some numerical constant $\frac12 \leq c \leq 1$.
  By the Diophantine property of $\omega$, for one $\xi_0$, there is at most one such $k_0\in\Z^d$ with $0<|k_0|\leq N_0$.
  (\ref{resonance_condition_0}) defines an interval ${\cal I}_{\la k_0\ra}\subset (-2, 2)$ of $E$.
  On ${\cal I}_{\la k_0\ra}$, a renormalization is necessary before the standard KAM procedure. More precisely, let
 $$H_{k_0,\, \tilde A_0}(\theta):=C_{\tilde A_0}\left(\begin{array}{cc}
                                   e^{{\rm i}\frac{\la k_0, \theta\ra}{2}} & 0 \\
                                   0 & e^{-{\rm i}\frac{\la k_0, \theta\ra}{2}}
                                 \end{array}\right)C_{\tilde A_0}^{-1}.$$
By a direct computation, we have
$$\tilde A_{\la k_0\ra}:=H_{k_0,\, \tilde A_0}(\theta+\omega)^{-1}\, \tilde A_0 \, H_{k_0,\, \tilde A_0}(\theta)=C_{\tilde A_0}\left(\begin{array}{cc}
                                   e^{{\rm i}(\tilde\alpha_0-\la k_0\ra)} & 0 \\
                                   0 & e^{-{\rm i}(\tilde\alpha_0-\la k_0\ra)}
                                 \end{array}\right)C_{\tilde A_0}^{-1}.$$
In view of Proposition 3 of \cite{HA}, we can see
\begin{equation}\label{Diophantine_new}
|(\xi_0-\la k_0\ra)-\la k\ra|\geq
\frac{\gamma}{|k|^\tau}-\frac{\varepsilon_0^\sigma}{|k|^\tau}\geq \frac{\gamma}{2|k|^\tau}, \quad \forall \, 0<|k|\leq N_0,
\end{equation}
and
$\tilde F_{\la k_0\ra}(\theta):=H_{k_0,\, \tilde A_0}(\theta+\omega)^{-1}\, \tilde F_0(\theta) \, H_{k_0,\, \tilde A_0}(\theta)$ is still bounded by $\varepsilon_0$.
  \item Case 2. If the condition (\ref{Diophantine_condition_0}) holds, let $k_0=0$ and the above procedure can be done trivially since $H_{k_0,\, \tilde A_0}= Id.$, $\tilde A_{\la k_0\ra}=\tilde A_0$, $\tilde F_{\la k_0\ra}=\tilde F_0$ and (\ref{Diophantine_condition_0}) implies (\ref{Diophantine_new}).
\end{itemize}

In both of the above cases, we can make a standard KAM procedure for $\tilde A_{\la k_0\ra}+\tilde F_{\la k_0\ra}(\theta)$ since the small divisor condition (\ref{Diophantine_new}) is always satisfied.
According to Proposition 6 of \cite{HA}, there exist
$$
\hat Z_{1}:(2\T)^d\rightarrow SL(2,\R),\quad \tilde A_{1}\in SL(2,\R), \quad \tilde F_{1}:\T^d\rightarrow gl(2,\R),
$$
such that
$\hat Z_{1}(\theta+\omega)^{-1}\left(\tilde A_0+\tilde F_0(\theta)\right)\hat Z_{1}(\theta)=\tilde A_{1}+\tilde F_{1}(\theta).$
The appearance of the intervals ${\cal I}_{\la k_0\ra}$, $0<|k_0|\leq N_0$, divides $[\inf\sigma(H),\, \sup\sigma(H)]$ into at most $|\ln\varepsilon_0|^{2d}$ connected components. $\hat Z_{1}$, $\tilde A_{1}$ and $\tilde F_{1}$ are ${\cal C}^2$ on each connected component, with, for $\nu=0,1,2$,
$$
|\partial^\nu(\hat Z_{1}-H_{k_0,\, \tilde A_0})|_{(2\T)^d}< \varepsilon_0^{\frac12}, \;\ |\partial^{\nu}(\tilde A_{1}-\tilde A_{\la k_0\ra})|< \varepsilon_0^{\frac23}, \;\ |\partial^{\nu} \tilde F_{1}|_{\T^d}\leq \varepsilon_{1}.
$$
Moreover, $\tilde A_1$ has two eigenvalues $e^{\pm{\rm i}\tilde\alpha_1}$ with $\xi_1:=\Re\tilde\alpha_1$ satisfying
$|\xi_1-(\xi_0-\la k_0\ra)|<\varepsilon_0^{\frac14}$ and $|\xi_1|<\frac32\varepsilon_0^\sigma$ on ${\cal I}_{\la k_0\ra}$, $k_0\neq 0$.

Since on each ${\cal I}_{\la k_0\ra}$, $|\xi_0-\la k_0\ra|< \frac{c\varepsilon_0^{\sigma}}{|k_0|^\tau}$, and $\xi_0$ is strictly increasing on $(-2,2)$, there is $E_*\in{\cal I}_{\la k_0\ra}$ such that $\xi_0(E_*)=\la k_0\ra$.
So $|{\rm tr}\tilde A_{\la k_0\ra}(E_*)|=2|\cos(\xi_0(E_*)-\la k_0\ra)|=2$.
As shown in \cite{E92}, after the standard KAM procedure which transforms $\tilde A_{\la k_0\ra}+\tilde F_{\la k_0\ra}$ to $\tilde A_1+\tilde F_1$, there maybe one subinterval ${\cal I}\subset{\cal I}_{\la k_0\ra}$(a neighbourhood of $E_*$), on which we have $|{\rm tr}\tilde A_1|>2$. Then on ${\cal I}$, $\xi_1\equiv0$. But on ${\cal I}_{\la k_0\ra}\setminus{\cal I}$, similar to the case of Corollary 6 of \cite{E92}, $\partial\xi_1>\frac13$, as the transversality of $\xi_0$.
Now, as a piecewise non-decreasing function, $\xi_1$ has the additional singularities at the edge of the interval ${\cal I}$'s.

\begin{itemize}
  \item [2)] The $(j+1)^{\rm th}-$step
\end{itemize}

Assume that we have arrived at $\tilde A_j+\tilde F_j(\theta)$, both of which are piecewise ${\cal C}^2$ with respect to $E$ on $[\inf\sigma(H),\, \sup\sigma(H)]$, with $\tilde A_j\in SL(2,\R)$, $\tilde F_j:\T^d\rightarrow gl(2,\R)$ satisfying
$$\tilde A_j=C_{\tilde A_j}\left(\begin{array}{cc}
e^{{\rm i}\tilde \alpha_j} & 0 \\
0 & e^{-{\rm i}\tilde \alpha_j}
\end{array}\right)C_{\tilde A_j}^{-1}, \quad |\partial^{\nu} \tilde F_{j}|_{(2\T)^d}\leq \varepsilon_{j}, \;\ \nu=0,1,2.$$
$\xi_j:=\Re\tilde\alpha_j$ is non-decreasing on each connected component where $\tilde A_j$ and $\tilde F_j$ are ${\cal C}^2$.

As in the first step, each connected component can be divided into at most $|\ln\varepsilon_j|^{2d}$ smaller components because of the appearance of intervals ${\cal I}_{\la k_j\ra}$, $0< |k_j|\leq N_j$, on which we have the resonances condition
\begin{equation}\label{resonance_condition_j}
|\xi_j -\la k_j \ra|< \frac{c\varepsilon_j^{\sigma}}{|k_j|^\tau}.
\end{equation}
As shown above, we can define, on ${\cal I}_{\la k_j\ra}$,
$$H_{k_j,\, \tilde A_j}(\theta):=C_{\tilde A_j}\left(\begin{array}{cc}
                                   e^{{\rm i}\frac{\la k_j, \theta\ra}{2}} & 0 \\
                                   0 & e^{-{\rm i}\frac{\la k_j, \theta\ra}{2}}
                                 \end{array}\right)C_{\tilde A_j}^{-1},$$
and, by a direct computation, we get
\begin{equation}\label{H_kj_Aj}
\tilde A_{\la k_j\ra}:=H_{k_j,\, \tilde A_j}(\theta+\omega)^{-1}\, \tilde A_j \, H_{k_j,\, \tilde A_j}(\theta)=C_{\tilde A_j}\left(\begin{array}{cc}
                                   e^{{\rm i}(\tilde\alpha_j-\la k_j\ra)} & 0 \\
                                   0 & e^{-{\rm i}(\tilde\alpha_j-\la k_j\ra)}
                                 \end{array}\right)C_{\tilde A_j}^{-1}.
\end{equation}
Outside ${\cal I}_{\la k_j\ra}$, we take $k_j=0$, and then $H_{k_j,\, \tilde A_j}\equiv Id.$ and $\tilde A_{\la k_j\ra}=\tilde A_j$.

With a similar procedure as above, we can find
$$
\hat Z_{j+1}:(2\T)^d\rightarrow SL(2,\R),\quad \tilde A_{j+1}\in SL(2,\R), \quad \tilde F_{j+1}:\T^d\rightarrow gl(2,\R),
$$
such that $\hat Z_{j+1}(\theta+\omega)^{-1}\left(\tilde A_j+\tilde F_j(\theta)\right)\hat Z_{j+1}(\theta)=\tilde A_{j+1}+\tilde F_{j+1}(\theta)$ with, for $\nu=0,1,2$,
\begin{equation}\label{appro_hatZ_tildeA_j+1}
|\partial^\nu(\hat Z_{j+1}-H_{k_j,\, \tilde A_j})|_{(2\T)^d}< \varepsilon_j^{\frac12}, \;\ |\partial^{\nu}(\tilde A_{j+1}-\tilde A_{\la k_j\ra})|< \varepsilon_j^{\frac23}, \;\ |\partial^{\nu} \tilde F_{j+1}|_{\T^d}\leq \varepsilon_{j+1}.
\end{equation}
$\tilde A_{j+1}$ has two eigenvalues $e^{\pm{\rm i}\tilde \alpha_{j+1}}$ with $\xi_{j+1}:=\Re\tilde \alpha_{j+1}$ satisfying
\begin{equation}\label{xi_j+1}
|\xi_{j+1}-(\xi_{j}-\la k_j\ra)|\leq \varepsilon_j^{\frac14}.
\end{equation}
Note that $\xi_{j+1}$ is a piecewise non-decreasing function of $E$. On every interval ${\cal I}_{\la k_j\ra}$, $k_j\neq0$, we have $|\xi_{j+1}|<\frac32\varepsilon_j^\sigma$, and there is a subinterval ${\cal I}\subset{\cal I}_{\la k_j\ra}$ on which $\xi_{j+1}\equiv0$, and on ${\cal I}_{\la k_j\ra}\setminus {\cal I}$, $\partial\xi_{j+1}> \frac13$.

At this moment, $[\inf\sigma(H), \sup\sigma(H)]$ is divided into
\begin{equation}\label{number_of_pieces}
\prod_{l=0}^j |\ln\varepsilon_l|^{2d}=|\ln\varepsilon_0|^{2(j+1)d} \prod_{l=0}^j(1+\sigma)^{2ld}\leq (1+\sigma)^{(j+1)^2 d}|\ln\varepsilon_0|^{2(j+1)d}\leq \frac{1}{10} |\ln\varepsilon_0|^{(j+1)^3 d}
\end{equation}
connected components, on which $\tilde A_{j+1}$ and $\tilde F_{j+1}$ are ${\cal C}^2$.
From the construction, we can see that each component is labelled with $\{k_l\}_{0\leq l \leq j}$, $|k_l|\leq N_l$.
Let $\tilde Z_{j+1}:=\prod_{l=j}^0\hat Z_{l+1}$.
In view of Proposition 3 of \cite{HA} and a direct computation with (\ref{appro_hatZ_tildeA_j+1}),
we estimate $\tilde A_{j+1}$ and $\tilde Z_{j+1}$ essentially in two cases.
\begin{itemize}
  \item On the component, with $k_l=0$ for any $0\leq l\leq j$, we have
  \begin{equation}\label{tilde_ZA_j+1_0}
   |\partial^\nu(\tilde Z_{j+1}-Id)|_{(2\T)^d}< \varepsilon_0^{\frac12}, \;\ |\partial^{\nu}(\tilde A_{j+1}-\tilde A_0)|< \varepsilon_0^{\frac23}, \quad \nu=0,1,2.
  \end{equation}
  \item On the component, where there exists $0\leq l\leq j$ such that
  $k_l\neq0$ and $k_{l'}=0$ for any $l< l'\leq j$, we have
  \begin{equation}\label{tilde_ZA_j+1}
\begin{array}{lll}
     |\tilde Z_{j+1}|_{(2\T)^d}\leq \varepsilon_l^{-\frac\sigma6}, & |\partial\tilde Z_{j+1}|_{(2\T)^d}\leq \varepsilon_l^{-\frac\sigma3},  & |\partial^2\tilde Z_{j+1}|_{(2\T)^d}\leq \varepsilon_l^{-\frac{\sigma}2},\\[1mm]
|\tilde A_{j+1}|\leq 5, & |\partial\tilde A_{j+1}|\leq N_l^{4\tau}, & |\partial^2\tilde A_{j+1}|\leq \varepsilon_l^{-\frac\sigma6}.
   \end{array}
\end{equation}
\end{itemize}

Let $\rho_{j+1}:=\xi_{j+1}+\sum_{l=0}^j\la k_l \ra$.
By Lemma 4 of \cite{HA}, for $\rho=\rho_{(\omega,\,  A_0+F_0)}$, we have $|\rho_{j+1}-\rho|_{\R}\leq \varepsilon_j^{\frac14}$. Moreover, by (\ref{xi_j+1}) and the resonance condition (\ref{resonance_condition_j}), we have
\begin{equation}\label{size_I_k_j}
\varepsilon_j^{\sigma(1+\frac\sigma2)}<\rho({\cal I}_{\la k_j\ra})\leq 10\varepsilon_j^{\sigma}.
\end{equation}

\begin{itemize}
  \item [3)] The limit state
\end{itemize}

As the iteration continues, we can finally get a sequence $\{k_j\}_{j\geq 0}$.
It is shown in Lemma 4 of \cite{HA} that, for a.e. $E\in\sigma(H)$, $k_j\neq0$ only for finite $j$'s.
So we define the sets
\begin{eqnarray*}
  \Sigma_{0} &:=& \{E\in\sigma(H): k_{l}=0 \; {\rm for \; any} \; l\geq 0\}, \\[1mm]
  \Sigma_{j+1} &:=& \{E\in\sigma(H): k_j\neq0 \;  {\rm and} \;  k_{l}=0 \; {\rm for} \; l\geq j+1\}, \;\ j\geq 0,
\end{eqnarray*}
with $|\sigma(H)\setminus\bigcup_{j\geq0}\Sigma_{j}|=0$. Obviously, $\Sigma_{j+1}$ is contained in the union of intervals ${\cal I}_{\la k_j\ra}$ obtained at the $j^{\rm th}-$step according to the resonance condition. So, by (\ref{number_of_pieces}) and (\ref{size_I_k_j}),
$|\rho\left(\Sigma_{j+1}\right)|\leq |\ln\varepsilon_0|^{(j+1)^3 d}\,  \varepsilon_{j}^{\sigma}$. Moreover, on $\Sigma_{j+1}$,
$$\left|\rho-\sum_{l\geq0}\la k_l\ra\right|\leq\left|\rho_{j+1}-\sum_{l\geq0}\la k_l\ra\right|+|\rho-\rho_{j+1}|
=|\xi_{j+1}|+|\rho-\rho_{j+1}|
\leq 2\varepsilon_{j}^{\sigma}.$$
Combining with the gap-labelling theorem, (2) is shown.

On $\Sigma=\bigcup_{j\geq0}\Sigma_{j}$, (\ref{appro_hatZ_tildeA_j+1}) implies the convergence of $\tilde Z_{j+1}=\prod_{l=j}^0\hat Z_{l+1}$, $\partial\tilde Z_{j+1}$ and $\tilde A_{j+1}$, $\partial\tilde A_{j+1}$ as $j\rightarrow\infty$.
Hence, for a.e. $E\in\sigma(H)$, we can define
$\tilde Z:=\lim_{j\rightarrow\infty}\tilde Z_j$ and $\tilde B=\lim_{j\rightarrow\infty}\tilde A_j\in SL(2,\R)$, such that
\begin{equation}\label{reducibility_of_Eliasson}
\tilde Z(\theta+n\omega)^{-1} \, (\tilde A_0+\tilde F_0(\theta)) \, \tilde Z(\theta)=\tilde B.
\end{equation}
This is exactly the reducibility obtained by the KAM scheme in \cite{E92} and \cite{HA}.

\smallskip

In addition, we define $\breve{Z}:=\lim_{j\rightarrow\infty}\partial\tilde Z_{j+1}$ and $\breve{B}:=\lim_{j\rightarrow\infty}\partial\tilde A_{j+1}$.
By (\ref{tilde_ZA_j+1_0}), (\ref{tilde_ZA_j+1}) and the definition of resonance sets $\Sigma_j$, we get
\begin{equation}\label{tilde_ZB}
\begin{array}{llll}
 |\tilde Z-Id.|_{\Sigma_0, (2\T)^d} \leq 2\varepsilon_0^{\frac12},& |\breve Z|_{\Sigma_0, (2\T)^d} \leq2 \varepsilon_0^{\frac12} ,&|\tilde B-A_0|_{\Sigma_0}\leq 2\varepsilon_0^{\frac23}, & |\breve B-\partial A_0|_{\Sigma_0}\leq 2\varepsilon_0^{\frac23} \\[1mm]
 |\tilde Z|_{\Sigma_{j+1}, (2\T)^d}\leq 2\varepsilon_{j}^{-\frac{\sigma}6}, &|\breve Z|_{\Sigma_{j+1}, (2\T)^d}\leq 2\varepsilon_{j}^{-\frac{\sigma}3}, &|\tilde B|_{\Sigma_{j+1}}\leq 6, & |\breve B|_{\Sigma_{j+1}}\leq 2 N_j^{4\tau}
\end{array}.
\end{equation}

\begin{itemize}
  \item [Part 2.] An additional transformation
\end{itemize}

On $\Sigma=\cup_{j\geq0}\Sigma_j$, the eigenvalues of $\tilde B=\left(\begin{array}{cc}
              \tilde B_{11} & \tilde B_{12} \\
              \tilde B_{21} & \tilde B_{22}
            \end{array}\right)$ is $e^{\pm{\rm i}\xi}$ with $\xi=\lim_{l\rightarrow \infty}\xi_l$ satisfying $\xi=\rho-\sum_{j\geq 0} \la k_j\ra$.
Let
\begin{equation}\label{additional_transformation}
H(\theta):= C_{\tilde B} \left(\begin{array}{cc}
                          \exp\{-\frac{\rm i}{2}\sum_{j\geq 0}\la k_j, \theta\ra\} & 0 \\[1mm]
                          0 & \exp\{\frac{\rm i}{2}\sum_{j\geq 0}\la k_j, \theta\ra\}
                        \end{array}
\right) C_{\tilde B}^{-1},
\end{equation}
with $C_{\tilde B}$ the matrix of normalized eigenvectors of $\tilde{B}$.
Then, with $Z:=\tilde Z\cdot H$ and $B:= C_{\tilde B} \left(\begin{array}{cc}
                          e^{i\rho} & 0 \\
                          0 & e^{-i\rho}
                        \end{array}
\right) C_{\tilde B}^{-1}$, it is easy to see that
$$
\tilde Z(\theta+\omega)\, \tilde B\,  \tilde Z^{-1}(\theta)=Z(\theta+\omega)\, B \, Z(\theta)^{-1}.
$$
So (1) is shown. Noting that
$C_{\tilde B}$ is just a normalization of $\left(\begin{array}{cc}
                      \tilde B_{12} & \tilde B_{12} \\
                      e^{{\rm i}\xi}-\tilde B_{11} &  e^{-{\rm i}\xi}-\tilde B_{11}
                    \end{array}\right)$,
by a direct computation, we get, on $\Sigma$,
\begin{eqnarray}
H(\theta)&=&\frac{\sin\frac{\sum_{j\geq0}\la k_j, \theta\ra}2}{\sin\xi}\left(\begin{array}{cc}
              \tilde B_{11} & \tilde B_{12} \\
              \tilde B_{21} & -\tilde B_{11}
            \end{array}\right) + \frac{\sin\left(\xi-\frac{\sum_{j\geq0}\la k_j, \theta\ra}2\right)}{\sin\xi}Id.,\label{additional_procedure_H}\\[1mm]
B&=&\frac{\sin\rho}{\sin\xi}\left(\begin{array}{cc}
              \tilde B_{11} & \tilde B_{12} \\
              \tilde B_{21} &  -\tilde B_{11}
            \end{array}\right) +\frac{1}{\sin\xi}\left(\begin{array}{cc}
                                         -\sin(\rho-\xi) & 0 \\
                                         0 & \sin(\rho+\xi)
                                       \end{array}
            \right).\label{additional_procedure_B}
\end{eqnarray}
In particular, on $\Sigma_0$, $H=Id.$, $B=\tilde B$ and hence $Z=\tilde Z$.
Recalling that $0<|\xi|< 2 \varepsilon_{j}^{\sigma}$ on $\Sigma_{j+1}$,
we have, by (\ref{tilde_ZB}),
\begin{equation}\label{ZB}
\left\{\begin{array}{ll}
| Z-Id.|_{\Sigma_0,\, (2\T)^d}, & |B- A_0|_{\Sigma_0}\leq 2\varepsilon_0^{\frac12}  \\[1mm]
|\sin^{2}\xi\cdot Z|_{\Sigma_{j+1},\, (2\T)^d}, & |\sin^{2}\xi\cdot B|_{\Sigma_{j+1}}\leq N_j \varepsilon_j^{\frac{5\sigma}{6}}
\end{array}\right. .
\end{equation}
\qed

\begin{Remark}[about the additional transformaion $H$]\label{rmq_addition_trans}
For the constant matrix $\tilde B$ in (\ref{reducibility_of_Eliasson}), its eigenvalues are $e^{\pm{\rm i}\xi}$ on $\Sigma$, with $\xi$ the renormalized rotation number.
According to the construction of $\xi_j$, it is piecewise non-decreasing and it is not uniquely determined(depending on the choice of coefficient $c$ in the resonance condition (\ref{resonance_condition_j})).
To apply the regularity(see Proposition \ref{proposition_rotation_number} in Subsection \ref{section_rotation_number}) and the uniqueness of the ``real" rotation number $\rho=\rho_{(\omega, A_0+F_0)}$, we need to conjugate $\tilde B$ to $B$ which has eigenvalues $e^{\pm{\rm i}\rho}$.
As shown in (\ref{additional_procedure_H}) and (\ref{additional_procedure_B}), this additional procedure brings us the singularities ``$\sim\frac{1}{\sin\xi}$" on $\Sigma_{j+1}$, $j\geq 1$, where $\xi$ is close to zero. Hence, on $\Sigma_{j+1}$, we need a smoothing factor $\sin^{2}\xi$ to cover the singularities and get better control on $Z$ and $B$, as shown in (\ref{ZB}).
\end{Remark}

\

Given $M\in\Z\setminus\{0\}$, with $J=J(M):=\min\left\{j\in\N: \; |M|\leq\varepsilon_{j}^{-\sigma}\right\}$, an approximation for the reducibility of quasi-periodic Schr\"odinger cocycle $(\omega, A_0+F_0)$ can be stated in the following way, which will be contributed to computing an integral on $[\inf\sigma(H), \sup\sigma(H)]$(see Subsection \ref{subsec_integral}).
\begin{Proposition}\label{propsana1}
Let $|V|_r=\varepsilon_0\leq \varepsilon_*$ be as in Proposition \ref{propsana}.
There is $$\Gamma^{(M)}=\bigcup_{j=0}^{J+1}\Gamma^{(M)}_j\subset[\inf\sigma(H), \sup\sigma(H)],$$
with $\{\Gamma^{(M)}_j\}_{j=0}^{J+1}$ mutually disjoint and $\Sigma_{j}\subset \Gamma^{(M)}_{j}$, satisfying
\begin{equation}\label{measure}
\sharp\left([\inf\sigma(H), \sup\sigma(H)]\setminus \Gamma^{(M)}\right)\leq |\ln\varepsilon_0|^{(J+1)^3 d}, \;\ \left|\rho\left(\Gamma^{(M)}_{j+1}\right)\right|\leq |\ln\varepsilon_0|^{(j+1)^3 d} \varepsilon_{j}^{\sigma},
\end{equation}
and $\left\{ \begin{array}{l}
                              A^{(M)}:\Gamma^{(M)}\rightarrow SL(2,\R) \;\ with \;\ two \;\ eigenvalues \;\ e^{\pm{\rm i}\alpha^{(M)}} \\[1mm]
                              Z^{(M)}:\Gamma^{(M)}\times(2\T)^d\rightarrow SL(2,\R) \;\ analytic \;\  on \;\ (2\T)^d
                            \end{array}
 \right.$, such that the following statements hold.\\
 \smallskip
\noindent {\bf (S1)} $|\Re\alpha^{(M)}-\rho|_{\Gamma^{(M)}}\leq \varepsilon_J^{\frac14}$ and for $0\leq j\leq J$, there is $k^{(M)}_j:\Gamma^{(M)}\rightarrow\Z^d$, constant on each connected component of $\Gamma^{(M)}$, such that
\begin{itemize}
\item [1.] $|k^{(M)}_{l}|_{\Gamma^{(M)}_{j}}=0$ if $l\geq j$,
\item [2.] $0<|k^{(M)}_j|\leq N_{j}$ on $\Gamma^{(M)}_{j+1}$ and $|\Re\alpha^{(M)} - \sum_{l=0}^{J}\la k^{(M)}_l\ra|_{\Gamma^{(M)}_{j+1}}\leq \frac32 \varepsilon_j^{\sigma}$, $0\leq j\leq J$.
\end{itemize}
{\bf (S2)} Let $\xi^{(M)}:=\Re\alpha^{(M)} - \sum_{l=0}^{J}\la k^{(M)}_l\ra$.
\begin{itemize}
\item On $\Gamma^{(M)}_{j+1}$, $0\leq j\leq J$, in each connected component,
      there is one and only one subinterval ${\cal I}$ such that
       $\xi^{(M)}=0$ on ${\cal I}$, and outside ${\cal I}$, $\sin\xi^{(M)}\neq0$ with
    \begin{equation}\label{esti_plat}
    \frac13<\partial\xi^{(M)}\leq N_j^{4\tau} |\sin\xi^{(M)}|^{-1}, \quad |\partial^2 \xi^{(M)}|\leq N_j^{8\tau}|\sin\xi^{(M)}|^{-3}.
    \end{equation}
\item On $\Gamma^{(M)}_{0}$, if $\sin\xi^{(M)}\neq 0$, we have $\partial\xi^{(M)}=-\frac{\partial {\rm tr}A^{(M)}}{2\sin\xi^{(M)}}> \frac13$. \footnote{Indeed, the only possibility that $\sin\xi^{(M)}=0$ on $\Gamma_0$ is on the intervals containing $\inf\sigma(H)$ and $\sup\sigma(H)$, as $\xi_0$ given in (\ref{xi_0}).}
\end{itemize}
{\bf (S3)} $|Z^{(M)}-Z|_{\Sigma_0,\, (2\T)^d}$, $|A^{(M)}- B|_{\Sigma_0} \leq \varepsilon_{J}^{\frac14}$, and for $0\leq j\leq J$,
$$|\sin\xi^{(M)}\, Z^{(M)}- \sin\xi\, Z|_{\Sigma_{j+1},\, (2\T)^d},\quad |\sin\xi^{(M)}\, A^{(M)}-\sin\xi\, B|_{\Sigma_{j+1}}\leq \varepsilon_{J}^{\frac14},$$
and for $\nu=0,1,2$,
\begin{equation}\label{sigma_m_0}
\left\{
\begin{array}{l}
\displaystyle |\partial^\nu (Z^{(M)}-Id.)|_{\Gamma^{(M)}_{0}, \, (2\T)^d},\; |\partial^\nu (A^{(M)}- A_0)|_{\Gamma^{(M)}_{0}}\leq \varepsilon_0^{\frac13}  \\[1mm]
\displaystyle |\partial^\nu Z^{(M)}|_{(2\T)^d}, \; |\partial^\nu A^{(M)}|\leq \frac{\varepsilon_j^{-\frac{\sigma}5}}{\sin^{1+2\nu}\xi^{(M)}} \  on \ \Gamma^{(M)}_{j+1}  \  if  \ \sin\xi^{(M)}\neq 0
\end{array}
\right. .
\end{equation}
{\bf (S4)} $\{E\in\partial\Gamma^{(M)}:M\rho(E)\notin\pi\Z\}\subset \partial\Gamma^{(M)}_{J+1}$.
     For any connected component $(E_*, E_{**})$ of $\Gamma^{(M)}_{J+1}$, we have
     $$\left|\left. \rho\right|_{(E_*, E_{**})}\right|\leq 2 \varepsilon_{J}^{\sigma(1+\frac\sigma2)}, \quad \varepsilon_J^{3\sigma(1+\sigma)}\leq E_{**}-E_{*}\leq \varepsilon_J^{\sigma(1+\frac{\sigma}3)}.$$
     Moreover, $k_j^{(M)}(E^-_*)=k_j^{(M)}(E^+_{**})$, $0\leq j\leq J$, and
     there is $0\leq j_* < J$ such that $E_*$, $E_{**}\in \partial\Gamma^{(M)}_{j_*}$, with
\begin{equation}\label{edge_point}
\left\{\begin{array}{llll}
\left|\left.(Z^{(M)}-Id.)\right|^{E^+_{**}}_{E^-_*}\right|_{(2\T)^d},   & \left|\left.( A^{(M)}-A_0)\right|^{E^+_{**}}_{E^-_*} \right|   &  \leq \displaystyle \frac{\varepsilon_0^{\frac13}}2(E_{**}-E_*), & j_*=0 \\[3mm]
\left|\left.\sin^4\xi^{(M)}\, Z^{(M)}\right|^{E^+_{**}}_{E^-_*} \right|_{(2\T)^d},   & \left|\left.\sin^4\xi^{(M)}\, A^{(M)}\right|^{E^+_{**}}_{E^-_*}\right|  &  \leq \displaystyle \frac{\varepsilon_{j_*-1}^{\frac{2\sigma}{3}}}2(E_{**}-E_*), & j_*\geq1
\end{array}\right.  .
\end{equation}
\end{Proposition}

\begin{Remark}[about construction of transformations]
$Z^{(M)}$, $A^{(M)}$ in Proposition \ref{propsana1} are constructed by KAM iteration as in the above proof.
They are just the above $\tilde Z_{J+1}$ and $\tilde A_{J+1}$, up to a renomalization which translates $\xi_{J+1}$(i.e., $\xi^{(M)}$) to $\rho_{J+1}$(i.e., $\Re\alpha^{(M)}$).
As mentioned in Remark \ref{rmq_addition_trans}, the construction of transformations $\{\hat Z_{j+1}\}$ is not uniquely determined in view of the above proof(depending on the coefficient $c$ in the resonance condition (\ref{resonance_condition_j})).
In particular, as shown in {\bf (S4)}, for any given non-zero integer $M$,
we can choose delicately the coefficient $c$, hence the endpoints of the ``resonance intervals" ${\cal I}_{\la k_j \ra}$, at the initial several steps,
such that $M\rho\in \pi\Z$ on these endpoints(since $\varepsilon_J^\sigma<\frac{1}{|M|}\leq\varepsilon^{\sigma}_{J-1}$, if $J>1$, $\rho({\cal I}_{\la k_j \ra})$ is adjustable within this range when $j< J$).
\end{Remark}

\begin{Remark} [about construction of resonance sets]
The mutually disjoint subsets $\{\Gamma^{(M)}_j\}_{0\leq j\leq J+1}$ given in Proposition \ref{propsana1} cover $[\inf\sigma(H), \sup\sigma(H)]$ up to finite points.
They divide the energies according to the extent of resonances.
As the iteration continues until the limit state, we can get the sequence of mutually disjointed subset $\{\Sigma_j\}_{j\geq 0}$ after excluding every gap in the spectrum.
\end{Remark}

\begin{Remark} [about the ``external variation"]
In {\bf (S4)}, we describe the size of the interval
${\cal I}_{\la k_J \ra}=(E_*, E_{**})$ obtained in the $J^{\rm th}-$step.
Besides the internal variation(the variation between $E_*^+$ and $E_{**}^-$), which is guaranteed by the ${\cal C}^2$ property,
the variation at the outer bounds of ${\cal I}_{\la k_J\ra}$, as shown in (\ref{edge_point}), is also needed for considering an integral on $[\inf\sigma(H),\sup\sigma(H)]$ in Subsection \ref{subsec_integral}.
The outer bounds of ${\cal I}_{\la k_J\ra}$ correspond to the non-resonance case, and one step before, they are both contained in one connected component, so the external variation can be obtained by the ${\cal C}^2$ property in the previous step.
Here the subscript $j_*$ represents the step when the last resonance and renormalization occurs(in particular, $j_*=0$ means is no resonance before the $(J+1)^{\rm th}-$step).
\end{Remark}

We shall give a proof of Proposition \ref{propsana1} in Appendix \ref{proofP2}.

\subsection{Application 1: absolutely continuous spectrum}\label{ac_spec}
\noindent

Eliasson\cite{E92} has shown the purely absolutely continuous spectrum for the continuous Schr\"odinger operator, based on the analysis on the
corresponding Schr\"odinger cocycle. But for the discrete operator, the purely absolutely continuous spectrum has not yet been explicitly proven.
In this subsection, a proof will be given based on some important estimates in \cite{HA}.

\begin{Proposition}\label{spec_ac}
With $|V|_r=\varepsilon_0\leq \varepsilon_*$ as in Proposition \ref{propsana}, we have, for any $\theta\in\T^d$, the spectrum of $H$ is purely absolutely continuous.
\end{Proposition}
We are going to prove the purely absolute continuity of the spectral measure $\mu_\theta=\mu^{e_{-1}}_\theta+\mu^{e_0}_\theta$ given in Subsection \ref{pre_op_cocycle}.
The main idea is the same with \cite{Avila}(see Subsection 3.8 of \cite{Avila}), which shows the absolutely continuous spectrum in one-frequency case.

\

Given $n\in \Z_+$, let
${\cal A}_n(E, \theta):=\prod_{j=n-1}^0(A_0(E)+F_0(\theta+j\omega))$.
We call that $(\omega, A_0+F_0)$ is {\bf bounded} if $\sup_{n\in\Z_+}|{\cal A}_n(E, \cdot)|_{(2\T)^d}<\infty$, and let ${\cal B}$ be the set of $E\in\sigma(H)$ such that $(\omega, A_0+F_0)$ is bounded.

Recalling the iteration process given in the previous proof,
for any $E\in \Sigma=\cup_{j\geq0}\Sigma_j$, any $l\geq1$, we have
$$
\tilde Z_{l}:(2\T)^d\rightarrow SL(2,\R),\quad \tilde A_{l}\in SL(2,\R), \quad \tilde F_{l}:(2\T)^d\rightarrow gl(2,\R),
$$
such that
$A_0+F_0(\theta)= \tilde Z_{l}(\theta+\omega) \left(\tilde A_{l}+\tilde F_{l}(\theta)\right) \tilde Z_{l}(\theta)^{-1}$ with $|\tilde F_{l}|_{\T^d}\leq \varepsilon_{l}$ and $|\tilde Z_l|_{(2\T)^d}\leq \varepsilon_{l-1}^{-\frac\sigma6}$.

\begin{Lemma}\label{bound_cocycle} For given $E\in\Sigma_j$, $j\geq 1$,
$\sup_{0\leq n \leq \varepsilon_j^{-4\sigma}}|{\cal A}_n|_{(2\T)^d} \leq \varepsilon_j^{-\frac{\sigma}{2}+\frac{\sigma^2}{4(1+\sigma)}}$.
\end{Lemma}
\proof For $E\in\Sigma_j$, $j\geq 1$, we know $\tilde A_{j}$ has eigenvalues $e^{\pm{\rm i}\xi_j}$ with $|\xi_j|\geq\frac{\gamma}{2N_j^{\tau}}$(note that the state $\tilde A_j+\tilde F_j$ means before the renormalization at the $(j+1)^{\rm th}-$step, so $\xi_j$ is close to some $\la k_j\ra$, $0<|k_j|<N_j$).
With $C_{\tilde A_{j}}$ the matrix of normalized eigenvectors of $\tilde A_j$, we have $|C_{\tilde A_j}| \leq 6$, and by Remark 3 of \cite{HA},
$$|C_{\tilde A_j}^{-1}|\ll \frac{2 \varepsilon_{j-1}^{-\frac\sigma6}}{|\xi_j|}\leq \frac{2N_j^{\tau}}{\gamma} \varepsilon_{j-1}^{-\frac\sigma6}.$$
Let $\hat F_j:=C^{-1}_{\tilde A_j}\tilde F_jC_{\tilde A_j}$, we have $|\hat F_j|_{\T^d}\leq \varepsilon_j^{1-\frac\sigma6}$, and
$$\tilde A_j+\tilde F_j(\theta)=C_{\tilde A_j} \left[ \left(\begin{array}{cc}
                                                    e^{{\rm i}\xi_j} & 0 \\[1mm]
                                                    0 & e^{-{\rm i}\xi_j}
                                                  \end{array}
   \right) + \hat F_j(\theta) \right]C_{\tilde A_j}^{-1}.$$
Then, for $0\leq n\leq \varepsilon_j^{-4\sigma}$,
\begin{eqnarray*}
{\cal A}_n(\theta)&=& \tilde Z_j(\theta+n\omega) \left(\tilde A_j+\tilde F_j(\theta+(n-1)\omega)\right) \cdots \left(\tilde A_j+\tilde F_j(\theta)\right) \tilde Z_j(\theta)^{-1}  \\
   &=& \tilde Z_j(\theta+n\omega) \cdot C_{\tilde A_j} \prod_{l=n-1}^0 \left[ \left(\begin{array}{cc}
                                                    e^{{\rm i}\xi_j} & 0 \\[1mm]
                                                    0 & e^{-{\rm i}\xi_j}
                                                  \end{array}
   \right) + \hat F_j(\theta+l\omega) \right] C_{\tilde A_j}^{-1} \cdot  \tilde Z_j(\theta)^{-1}.
\end{eqnarray*}
So $|{\cal A}_n|_{(2\T)^d} \leq \frac{2N_j^{\tau}}{\gamma} \varepsilon_{j-1}^{-\frac\sigma2}\leq\varepsilon_j^{-\frac{\sigma}{2}+\frac{\sigma^2}{4(1+\sigma)}}$.\qed

We also have the following lemmas, which is generalized from the case $d=1$. Since the proof can be directly translated, we do not present them precisely.


\begin{Lemma}[Lemma 2.5 of \cite{Avila}]\label{lemma2.5_of_avila}
There is a universal constant $C>0$, independent of $\theta$, such that
for every $\theta\in \T^d$, $\mu_{\theta}(E-\epsilon, E+\epsilon)\leq C\epsilon \sup_{0 \leq n \leq C\epsilon^{-1}} |{\cal A}_n(E)|^2_{(2\T)^d}$.
\end{Lemma}

\begin{Lemma}[Theorem 2.4 of \cite{Avila}]\label{theorem2.4_of_avila}
For every $\theta\in \T^d$, $\mu_{\theta}|_{\cal B}$ is absolutely continuous.
\end{Lemma}


\noindent{\bf Proof of Proposition \ref{spec_ac}:} Fix $\theta\in\T^d$ and we do not present it explicitly.
By Lemma \ref{theorem2.4_of_avila}, it is enough to show that $\mu(\sigma(H)\setminus {\cal B})=0$.
Let ${\cal R}$ be the set of $E\in\R$ such that $(\omega, A_0+F_0)$ is reducible.
Notice that ${\cal R}\setminus{\cal B}$ contains only $E$ such that $(\omega, A_0+F_0)$ is analytically reducible to
parabolic. It follows that ${\cal R}\setminus{\cal B}$ is countable: indeed for any $E\in{\cal R}\setminus{\cal B}$, there exists $k\in\Z^d$ such that $\rho(\omega, A_0+F_0) = \la k\ra$.
If $E\in\cal R$, any nonzero solution of $Hq=Eq$
satisfies $\inf_{n\in\Z} \{|q_n|^2+|q_{n+1}|^2\} > 0$. In particular, there are no eigenvalues in ${\cal R}$, and
$\mu({\cal R}\setminus{\cal B}) = 0$.
Thus, we only need to prove that $\mu(\sigma(H)\setminus{\cal R}) = 0$.

Let $K_m\subset \sigma(H)$, $m\geq 0$, be the set of $E$ such that the rotation number $\rho$ satisfies
$$|\rho-\la k \ra|\leq \frac{\varepsilon_m^\sigma}{|k|^\tau} \;   {\rm  for \; some } \;  N_m < |k| \leq N_{m+1}.$$
Obviously, $K_m\subset \overline{\cup_{j\geq m+1}\Sigma_j}$.
In view of Proposition \ref{propsana} and Remark \ref{rmq_Prop1}, if the resonance stops at one finite step, the cocycle $(\omega,\, A_0+F_0)$ is reducible. So we have $\sigma(H)\setminus{\cal R}\subset \limsup K_m$.
By Borel-Cantelli lemma, $\sum_{m\geq 0}\mu(\overline{K}_m)< \infty$ implies that $\mu(\sigma(H)\setminus{\cal R}) = 0$.

Now we are going to show that $\sum_{m\geq 0}\mu(\overline{K}_m)< \infty$.
For every $E\in\Sigma_j$, $j\geq m+1$, we know that $|\rho(E)-\la k\ra|\leq 2\varepsilon_j^{\sigma}$ for some $|k|\leq N_j$.
This shows that $\Sigma_j$ can be covered by $10N^d_j$ intervals $T_s$ of length $2\varepsilon_j^{\sigma}$.
By Lemma \ref{bound_cocycle} and Lemma \ref{lemma2.5_of_avila}, for any $s$,
$$\mu(T_s)\leq C\cdot 2\varepsilon_j^{\sigma}\sup_{0 \leq n \leq \frac{C}2\varepsilon_j^{-\sigma}} |{\cal A}_n(E)|^2_{(2\T)^d}\leq 2C \varepsilon_j^{\frac{\sigma^2}{2(1+\sigma)}}.$$
Then $\mu(\Sigma_j)\leq \varepsilon_j^{\frac{\sigma^2}{3}}$. So
$\mu(K_m)\leq\sum_{j\geq m+1} \mu(\Sigma_j)<\varepsilon_{m+1}^{\frac{\sigma^2}{5}}$,
which gives $\sum_{m\geq 0}\mu(\overline{K}_m)< \infty$. \qed

\subsection{Application 2: regularity and transversality of rotation number}\label{section_rotation_number}
\noindent

For the rotation number $\rho=\rho_{(\omega,\, A_0+F_0)}$,
we also have the following further results, which come with the analysis on the reducibility of Schr\"odinger cocycle.
\begin{Proposition}\label{proposition_rotation_number}
With $|V|_r=\varepsilon_0\leq \varepsilon_*$ as in Proposition \ref{propsana}, we have
\begin{enumerate}
  \item $\rho=\rho_{(\omega,\, A_0+F_0)}$ is $\frac12-$H\"{o}lder continuous, i.e., there is a numerical constant $c>0$, such that for any given $E_1$, $E_2\in\R$,
  $$|\rho(E_1)-\rho(E_2)|<c |E_1-E_2|^{\frac12}.$$
  \item $\rho=\rho_{(\omega,\, A_0+F_0)}$ is absolutely continuous on $\R$, i.e., given finite intervals $\{{\cal I}_j\}_j$ on $\R$, for any $\eta>0$, there exists $\delta=\delta(\eta)>0$, such that if $\sum_j|{\cal I}_j| < \delta$ then $\sum_j \left|\left.\rho\right|_{{\cal I}_j}\right| <\eta$.
\end{enumerate}
\end{Proposition}
\proof Recalling that $A_0(E)=\left(\begin{array}{cc}
            -E & -1 \\
            1 & 0
          \end{array}
\right)$ and $F_0(\theta)=\left(\begin{array}{cc}
            V(\theta) & 0 \\
            0 & 0
          \end{array}
\right)$,
the H\"{o}lder continuity and absolute continuity are obtained as direct corollaries of Theorem 2 in \cite{HA} and Theorem 1 in \cite{HA2014} respectively. \qed

\begin{Proposition}\label{proposition_rotation_number}
With $|V|_r=\varepsilon_0\leq \varepsilon_*$ as in Proposition \ref{propsana}, we have
\begin{equation}\label{transversality_rho}
(2\sin\rho)^{-1}<\partial\rho<\infty \ for \ a.e. \ E\in\sigma(H).
\end{equation}
\end{Proposition}
\proof According to Proposition \ref{spec_ac} in the previous subsection, if $|V|_r\leq \varepsilon_*$, then the spectrum of $H_\theta$ is purely absolutely continuous for any $\theta\in\T^d$.
As the well-known result of Kotani theory, $L(E)=0$ for a.e. $E\in\sigma(H)$.
In view of Theorem 1.4 of \cite{DS}, we get the conclusion.\qed

\begin{Remark}
$\rho$ is non-decreasing and, in particular, constant outside the spectrum.
The transversality (\ref{transversality_rho}) of $\rho$ is related to the reducibility. More precisely, for the constant $B\in SL(2,\R)$ in Proposition \ref{propsana}, we have ${\rm tr}B=2\cos\rho$. Then $\partial\rho$ can be written(formally) as ``$-\frac{\partial{\rm tr}B}{2\sin\rho}$", which is similar to $\partial\xi_0$ in (\ref{partial_xi_0}) and {\bf (S2)} in Proposition \ref{propsana1}.
\end{Remark}

From now on, for convenience, we assume that (\ref{transversality_rho}) is satisfied on the full-measure subset $\Sigma$ of $\sigma(H)$ given in Proposition \ref{propsana}.

\subsection{Application 3: construction of Bloch-waves}\label{subsec_bloch}
\noindent

In general, the {\bf Bloch-wave} of a self-adjoint operator on $\ell^2(\Z)$ means the generalized eigenvector $\psi$, of the form
$\psi_n=e^{{\rm i}n\varrho} h(x+n\tilde\alpha)$, with $\varrho$, $\tilde\alpha$ some real numbers, and $h$ a periodic function of $x\in\R$.
Here $\varrho$ is called the {\bf Floquet exponent}, and its imaginary part is called the Lyapunov exponent.
In particular, if we consider the Schr\"odinger operator $H$, this definition of Lyapunov exponent is equivalent to that one given in Subsection \ref{pre_op_cocycle} for Schr\"odinger cocycle.

Back to Proposition \ref{propsana}, we can construct Bloch-waves of Schr\"odinger operator $H$ on
$\Sigma$.
More precisely, for the Schr\"odinger operator $H=H_\theta$, by the matrices
$Z=\left(\begin{array}{cc}
Z_{11} & Z_{12} \\
Z_{21} & Z_{22}
\end{array}\right)$ and
$B=\left(\begin{array}{cc}
B_{11} & B_{12} \\
B_{21} & B_{22}
\end{array}\right)$ given in Proposition \ref{propsana},
we can see $(\tilde\psi_n)_{n}=(e^{{\rm i}n\rho}\tilde f_n(\theta))_{n}$ is a solution of the equation $Hq=Eq$ for $E\in\Sigma$, with $\tilde f_n:\Sigma\times (2\T)^d\rightarrow \C$ given by
$$\tilde f_n(\theta):=\left[Z_{11}(\theta-\omega+n\omega)B_{12}-Z_{12}(\theta-\omega+n\omega)B_{11}\right]e^{-{\rm i}\rho}+Z_{12}(\theta-\omega+n\omega).$$
Indeed, by noting that $\left(\begin{array}{c}
B_{12} \\[1mm]
e^{{\rm i}\rho}-B_{11}
\end{array}
\right)$ is an eigenvector of $B$ corresponding to the eigenvalue $e^{{\rm i}\rho}$,
with $\left(\begin{array}{c}
\tilde\psi_{1} \\[1mm]
\tilde\psi_{0}
\end{array}
\right)=Z(\theta)\left(\begin{array}{c}
B_{12} \\[1mm]
e^{{\rm i}\rho}-B_{11}
\end{array}
\right)$, we get the generalized eigenvector
$$\left(\begin{array}{c}
\tilde\psi_{n+1} \\[1mm]
\tilde\psi_{n}
\end{array}
\right)=Z(\theta+n\omega) \, B^{n} \, Z(\theta)^{-1} \left(\begin{array}{c}
\tilde\psi_{1} \\[1mm]
\tilde\psi_{0}
\end{array}
\right)=e^{{\rm i}n\rho} \, Z(\theta+n\omega) \left(\begin{array}{c}
B_{12} \\[1mm]
e^{{\rm i}\rho}-B_{11}
\end{array}
\right).$$
Hence, we can also get the Bloch-wave
$$
\psi=(e^{{\rm i}n\rho}f_n)_{n\in\Z} \;\ {\rm with} \;\ f_n=\left\{\begin{array}{ll}
 \tilde f_n, & E\in\Sigma_0 \\[1mm]
\tilde f_n\sin^5\xi, & E\in\Sigma_{j+1} , \; j\geq0
\end{array} \right. .
$$

\begin{Remark} The Bloch-wave $(\tilde\psi_n)_{n\in\Z}$ depends on the energy $E\in\Sigma$.
Recall (\ref{limit_state_whitney}) and Remark \ref{rmq_Prop1}. On $\Sigma_0$, the large part of the spectrum, $(\tilde\psi_n)_{n}$ has nice estimates.
In contrast, it has some singularities ``$\sim\frac{1}{\sin\xi}$" on $\Sigma_{j+1}$, $j\geq 0$, whose union forms a small part of the spectrum.
So we add a smoothing factor $\sin^5\xi$, just on this small part to cover the singularities.
\end{Remark}

Based on the Bloch-wave $\psi$, we can introduce the ingredients of the modified spectral transformation for the Schr\"odinger operator(see Subsection \ref{modified_spec_trans}).
Let ${\cal K}_n:={\Im}(e^{{\rm i}n\rho}f_n \bar f_0)$ and
${\cal J}_n :={\Re}(e^{{\rm i}n\rho}f_n \bar f_0)$ on $\Sigma$ and ${\cal K}_n|_{\R\setminus \Sigma}={\cal J}_n|_{\R\setminus \Sigma}:=0$. By a direct calculation, we see
$$
e^{{\rm i}n\rho}f_n \bar f_0  =\sum_{n_\Delta= n, n\pm 1} \beta_{n,n_\Delta}e^{{\rm i}n_\Delta\rho},
$$
where $\beta_{n,n_\Delta}:\Sigma\times(2\T)^d\rightarrow \R$, analytic on $(2\T)^d$ and ${\cal C}^1_W$ on each $\Sigma_j$, $j\geq0$, is given by
$\beta_{n,n_\Delta}=\left\{\begin{array}{ll}
\tilde \beta_{n,n_\Delta}, & E\in\Sigma_0 \\[1mm]
     \tilde \beta_{n,n_\Delta}\sin^{10}\xi,&  E\in\Sigma_{j+1},\;\ j\geq0
  \end{array}\right.$,
with
\begin{eqnarray*}
\tilde \beta_{n,n}(\theta) &:=&Z_{12}(\theta-\omega+n\omega)Z_{12}(\theta-\omega)(1+B_{11}^2)+ Z_{11}(\theta-\omega+n\omega) Z_{11}(\theta-\omega) B_{12}^2
\nonumber\\[1mm]
       & &-\left[Z_{11}(\theta-\omega+n\omega) Z_{12}(\theta-\omega)+Z_{11}(\theta-\omega) Z_{12}(\theta-\omega+n\omega)\right]B_{11} B_{12},\\[1mm]
\tilde \beta_{n,n+1}(\theta) &:=&Z_{11}(\theta-\omega) Z_{12}(\theta-\omega+n\omega) B_{12}- Z_{12}(\theta-\omega+n\omega)Z_{12}(\theta-\omega)B_{11} ,\\[1mm]
\tilde \beta_{n,n-1}(\theta) &:=&Z_{11}(\theta-\omega+n\omega) Z_{12}(\theta-\omega) B_{12}-  Z_{12}(\theta-\omega+n\omega)Z_{12}(\theta-\omega)B_{11}.
\end{eqnarray*}
Then ${\cal K}_n=\sum_{n_\Delta}\beta_{n, n_\Delta} \sin n_\Delta\rho$,
${\cal J}_n=\sum_{n_\Delta}\beta_{n, n_\Delta} \cos n_\Delta\rho$.
In particular, $\beta_{0,1}=\beta_{0,-1}$, so
\begin{equation}\label{K0J0}
{\cal K}_0=0,\quad {\cal J}_0=\beta_{0,0}+2\beta_{0,1} \cos \rho.
\end{equation}
According to (\ref{limit_state_whitney}) and the fact that $|\xi|_{\Sigma_{j+1}}\leq 2\varepsilon_j^{\sigma}$, $j\geq 0$, it is obvious that
\begin{equation}\label{esti_abc_sigma_avant}
|\beta_{n,n_\Delta}-\delta_{n,n_\Delta}|_{{\cal C}^1_W(\Sigma_0),\, (2\T)^d}\leq \varepsilon_0^{\frac14},\quad |\beta_{n,n_\Delta}|_{{\cal C}^1_W(\Sigma_{j+1}),\, (2\T)^d}\leq \varepsilon_j^{\sigma},\;\  j \geq 0.
\end{equation}
Hence, for any $E\in\R$, $({\cal K}_n)_n$, $({\cal J}_n)_n\in\ell^\infty(\Z)$ with the $\ell^\infty-$norms bounded by $2$.

We have the following property about the coefficients $\beta_{n, n_\Delta}$.
\begin{Lemma}\label{lemma_nn}
For $m,n\in\Z$,
$\displaystyle \left|\int_{\Sigma} \beta_{m, m_\Delta} \beta_{n, n_\Delta}\, \partial\rho\, dE- \delta_{m, m_\Delta }\delta_{n, n_\Delta}\pi\right|_{(2\T)^d}\leq \varepsilon_0^{\frac{\sigma}{3}}$.
\end{Lemma}
\proof By (\ref{esti_abc_sigma_avant}), we can get for all $m, n \in\Z$,
$$\left|\beta_{m, m_\Delta} \beta_{n, n_\Delta}-\delta_{m, m_\Delta } \delta_{n, n_\Delta }\right|_{\Sigma_0,\, (2\T)^d}\leq 3\varepsilon_0^{\frac14};\quad |\beta_{m, m_\Delta} \beta_{n, n_\Delta}|_{\Sigma_{j+1},\, (2\T)^d}\leq \varepsilon_j^{2\sigma},\;\ j \geq 0.$$
Then, recalling that $|\rho\left(\Sigma_{j+1}\right)|\leq |\ln\varepsilon_0|^{(j+1)^3 d} \, \varepsilon_{j}^{\sigma}$,
 we have $$\left|\int_{\Sigma_j} \left(\beta_{m, m_\Delta} \beta_{n, n_\Delta}-\delta_{m, m_\Delta} \delta_{n, n_\Delta}\right)\, \partial\rho\,dE\right|_{(2\T)^d}\leq
\left\{ \begin{array}{cc}
          3\pi\varepsilon_0^{\frac14}, & j=0 \\[1mm]
          (1+\varepsilon_{j-1}^{2\sigma})\cdot|\ln\varepsilon_0|^{j^3 d} \, \varepsilon_{j-1}^{\sigma}, &  j\geq 1
        \end{array}
\right. .$$
Therefore,
$\displaystyle \left|\int_{\Sigma} \beta_{m, m_\Delta} \beta_{n, n_\Delta}\, \partial\rho\,dE-\delta_{m, m_\Delta} \delta_{n, n_\Delta}\pi\right|_{(2\T)^d}\leq\varepsilon_0^{\frac\sigma3}$.
\qed

With $Z$ and $B$ replaced by $Z^{(M)}$ and $A^{(M)}$ given in Proposition \ref{propsana1} respectively,
we can get $\tilde\beta_{n,n_\Delta}^{(M)}:\Gamma^{(M)}\times (2\T)^d\rightarrow \R$ in the same way as $\tilde\beta_{n,n_\Delta}$, and
$$\beta_{n,n_\Delta}^{(M)}=\left\{\begin{array}{ll}
\tilde \beta_{n,n_\Delta}^{(M)}, & E\in\Gamma_0^{(M)} \\[1mm]
\tilde \beta_{n,n_\Delta}^{(M)}\sin^{10}\xi^{(M)},&  E\in\Gamma_{j+1}^{(M)},\;\ 0\leq j\leq J
  \end{array}\right. .$$
Then $\tilde \beta_{n,n_\Delta}^{(M)}$ is ${\cal C}^2$ on each connected component of $\Gamma^{(M)}$.

\begin{Lemma}\label{coefficient_abc}
For every $n\in\Z$,
$$\left\{ \begin{array}{ll}
            |\partial^\nu(\beta^{(M)}_{n,n_\Delta} -\delta_{n, n_\Delta })|_{\Gamma^{(M)}_0,\, (2\T)^d}\leq \varepsilon_{0}^{\frac14}, & \\[1mm]
            |\partial^\nu \beta_{n,n_\Delta}^{(M)}|_{(2\T)^d} \leq \varepsilon_j^{\frac\sigma6}|\sin\xi^{(M)}|^{5-2\nu} \;\ on \;\ \Gamma^{(M)}_{j+1}, & 0\leq j\leq J
          \end{array}
 \right. ,\quad \nu=0,1,2,$$
and for each connected component $(E_*, E_{**})\subset \Gamma^{(M)}_{J+1}$,
 $ \left| \left.\beta^{(M)}_{n, n_\Delta}\right|^{E^+_{**}}_{E^-_*} \right|_{\Gamma^{(M)}_{J+1},\, (2\T)^d} \leq \varepsilon_{J}^{\sigma(1+\frac\sigma4)}$.
\end{Lemma}

\proof We only prove the statements for $\beta^{(M)}_{n, n}$, with that of $\beta^{(M)}_{n, n+1}$ and $\beta^{(M)}_{n, n-1}$ similar.

On $\Gamma^{(M)}_0$, $\beta^{(M)}_{n, n}=\tilde \beta^{(M)}_{n, n}$ equals to
\begin{eqnarray}
& &Z^{(M)}_{12}(\theta-\omega+n\omega)Z^{(M)}_{12}(\theta-\omega)\left[1+(A^{(M)}_{11})^2\right]+\, Z^{(M)}_{11}(\theta-\omega+n\omega) Z^{(M)}_{11}(\theta-\omega) (A^{(M)}_{12})^2
\nonumber\\[1mm]
& &-\left[Z^{(M)}_{11}(\theta-\omega+n\omega) Z^{(M)}_{12}(\theta-\omega)+Z^{(M)}_{11}(\theta-\omega) Z^{(M)}_{12}(\theta-\omega+n\omega)\right]A^{(M)}_{11} A^{(M)}_{12}.\label{tilde_an_m}
\end{eqnarray}
Then, in view of (\ref{sigma_m_0}), $|\partial^\nu(\beta^{(M)}_{n, n} - 1)|_{(2\T)^d}\leq \varepsilon_{0}^{\frac14}$ is evident.

On $\Gamma^{(M)}_{j+1}$, $0\leq j \leq J$, $\beta^{(M)}_{n, n}=\tilde \beta^{(M)}_{n, n}\sin^{10} \xi^{(M)}$.
In each connected component of $\Gamma^{(M)}_{j+1}$, according to {\bf (S2)}, $\beta^{(M)}_{n, n}=0$ on its subinterval ${\cal I}$ where $\xi^{(M)}=0$.
Outside ${\cal I}$, $0<|\sin\xi^{(M)}|<2\varepsilon_j^\sigma$, then
by (\ref{esti_plat}), (\ref{sigma_m_0}) and (\ref{tilde_an_m}), we have, for $\nu=0,1,2$,
$$|\partial^{\nu}\tilde \beta^{(M)}_{n, n}|_{(2\T)^d}\leq10\varepsilon_j^{-\frac{4\sigma}{5}}|\sin\xi^{(M)}|^{-(4+2\nu)},\quad
|\partial^{\nu} \sin^{10}\xi^{(M)}|\leq\frac{1}{10}\varepsilon_j^{\frac{29}{30}\sigma}|\sin\xi^{(M)}|^{9-2\nu}. $$
Hence, combining the estimates above, $|\partial^\nu \beta^{(M)}_{n, n}|_{(2\T)^d} \leq \varepsilon_j^{\frac\sigma6}|\sin\xi^{(M)}|^{5-2\nu}$ on $\Gamma^{(M)}_{j+1}$.

\smallskip

For the connected component $(E_*, E_{**})\subset \Gamma^{(M)}_{J+1}$, according to {\bf (S4)}, there is $0\leq j_*\leq J$, such that $E^-_{*}, E^+_{**}\in \partial\Gamma^{(M)}_{j_*}$. By (\ref{sigma_m_0}) and (\ref{edge_point}), and the fact that $E_{**}-E_*\leq\varepsilon_J^{\sigma(1+\frac{\sigma}3)}$,
\begin{itemize}
  \item if $j_*=0$,
$\left|\left.\beta^{(M)}_{n, n}\right|^{E^+_{**}}_{E^-_*}\right|_{(2\T)^d}
=\left|\left.\tilde \beta^{(M)}_{n, n}\right|^{E^+_{**}}_{E^-_*} \right|_{(2\T)^d}
\leq 10\, (E_{**}-E_*)\leq \varepsilon_J^{\sigma(1+\frac\sigma4)}$;
  \item if $j_* \geq 1$, then for $\beta^{(M)}_{n, n}=\tilde \beta^{(M)}_{n, n}\sin^{10}\xi^{(M)}$, $\left| \left.\tilde \beta^{(M)}_{n, n}\sin^{10}\xi^{(M)}\right|^{E^+_{**}}_{E^-_*} \right|_{(2\T)^d}$ can be bounded by terms like
$$20 \left| \left. \sin^{4}\xi^{(M)}\cdot Z^{(M)}\right|^{E^+_{**}}_{E^-_*} \right|_{(2\T)^d}\cdot\left| \sin^2\xi^{(M)}\cdot Z^{(M)} \right|^3_{\Gamma_{j_*}^{(M)},(2\T)^d}\leq \varepsilon_J^{\sigma(1+\frac\sigma4)}.$$\qed
\end{itemize}

Moreover, by {\bf (S3)}, it is obvious that
\begin{equation}\label{error_beta_n}
|\beta_{n, n_\Delta}-\beta^{(M)}_{n, n_\Delta}|_{\Sigma_j,\, (2\T)^d}\leq 10 \varepsilon_J^{\frac14}, \quad 0\leq j\leq J+1.
\end{equation}

\section{Proof of ballistic motion}\label{Proof_of_QP}

\subsection{An integral on $[\inf\sigma(H), \sup\sigma(H)]$ }\label{subsec_integral}
\noindent

Recall that in Proposition \ref{propsana1}, we have divided the interval $[\inf\sigma(H), \sup\sigma(H)]$ into $J(M)+2$ parts for some given $M\in\Z\setminus\{0\}$, up to a subset of finite points.
With this division, we can estimate the following integral, which will be applied in analyzing the modified spectral transformation.

\begin{Lemma}\label{hmn_integral}
Assume that $h$ is ${\cal C}^2$ on each connected component of $\Gamma^{(M)}$ given in Proposition \ref{propsana1}, satisfying
 \begin{itemize}
   \item[\rm (c1)]
   \begin{itemize}
     \item $|h|_{\Gamma^{(M)}_0}\leq 2$, $|\partial h|_{\Gamma^{(M)}_0}$, $|\partial^2 h|_{\Gamma^{(M)}_0}\leq \varepsilon_0^\frac16$,
     \item $|\partial^{\nu} h| \leq \varepsilon_j^{\frac\sigma3}|\sin\xi^{(M)}|^{5-2\nu}$ on $\Gamma^{(M)}_{j+1}$, $0\leq j\leq J$, for $\nu=0,1,2$.
   \end{itemize}
   \item[\rm (c2)] For any connected component $(E_*, E_{**})\subset \Gamma^{(M)}_{J+1}$, $\left|\left. h\right|^{E^+_{**}}_{E^-_*}\right|\leq \varepsilon_{J}^{\sigma(1+\frac\sigma6)}$.
 \end{itemize}
Then $\displaystyle \left|\int_{\inf\sigma(H)}^{\sup\sigma(H)} h\, \cos M \rho   \cdot \partial\rho\, dE\right|\leq
\frac{\varepsilon_0^{\frac{\sigma^2}{6}}}{|M|^{1+\frac{\sigma}{6}}}$.
\end{Lemma}

\proof The integral above is the sum of integrals over the connected component $(E_*,\, E_{**})\subset \Gamma^{(M)}$.
In view of Proposition \ref{proposition_rotation_number}, $\rho$ is absolutely continuous. So, by applying integration by parts on each connected component,
$$\int_{\inf\sigma(H)}^{\sup\sigma(H)} h\,  \cos M \rho   \cdot \partial\rho\, dE= \frac{1}{M}\sum_{(E_*, \, E_{**})\subset \Gamma^{(M)}\atop{\rm connected \ component}}\left(\left. h\, \sin M \rho\right|_{(E_*, E_{**})}-  \int_{E_*}^{E_{**}} (\partial h)\sin M \rho\,  dE\right).$$

Since {\bf (S4)} implies that $\sin M\rho(E)=0$ if $E\in\partial\Gamma^{(M)}\setminus\partial\Gamma_{J+1}^{(M)}$, we can see
\begin{equation}\label{sum_of_integral}
\sum_{(E_*, E_{**})\subset \Gamma^{(M)}\atop{\rm connected \ component}}\left. h\, \sin M \rho\right|_{(E_*, E_{**})} = \sum_{(E_*, E_{**})\subset\Gamma^{(M)}_{J+1}\atop{\rm connected \ component}} \left(
\left. h\, \sin M \rho\right|_{(E_*, E_{**})} -  \left. h\, \sin M \rho\right|_{E^-_*}^{E^+_{**}}\right) .
\end{equation}
Then, by (c1), (c2) and the fact $\left|\left. \rho\right|_{(E_*, E_{**})}\right|\leq 2 \varepsilon_{J}^{\sigma(1+\frac\sigma2)}$,
\begin{eqnarray*}
\left|\left. h\, \sin M\rho\right|^{E^+_{**}}_{E^-_*}\right| &\leq& \left|\sin M\rho\left(E_{**}\right)\right|\cdot\left|\left. h\right|^{E^+_{**}}_{E^-_*}\right|+\left|h\left(E^-_*\right)\right|\cdot\left|\left.\sin M\rho\right|_{(E_*, E_{**})}\right|\\
&\leq& \varepsilon_{J}^{\sigma(1+\frac\sigma6)}+2|M|\cdot 2 \varepsilon_{J}^{\sigma(1+\frac\sigma2)}\\
&\leq&5\varepsilon_J^{\frac{\sigma^2}2},
\end{eqnarray*}
and similarly $\left|\left. h\, \sin M \rho\right|_{(E_*, E_{**})}\right|\leq 5\varepsilon_J^{\frac{\sigma^2}2}$.
Recalling that there are at most $|\ln\varepsilon_0|^{(J+1)^3 d}$ connected components in $[\inf\sigma(H), \sup\sigma(H)]$, we get
\begin{equation}\label{esti_simple}
\left|\frac{1}{M}\sum_{(E_*, E_{**})\subset\Gamma^{(M)}\atop{\rm connected}}\left. h\, \sin M \rho\right|_{(E_*, E_{**})}\right|\leq \frac{|\ln\varepsilon_0|^{(J+1)^3d}}{|M|} \cdot 10\varepsilon_J^{\frac{\sigma^2}2}\leq \frac{\varepsilon_J^{\frac{\sigma^2}3}}{2|M|}\leq \frac{\varepsilon_J^{\frac{\sigma^2}6}}{2|M|^{1+\frac\sigma6}}.
\end{equation}

For the remaining part of integral, we consider $\int_{E_*}^{E_{**}} \, (\partial h)\sin  M \rho^{(M)}\,  dE$ instead, with $\rho^{(M)}:=\Re\alpha^{(M)}$.
Indeed, in view of {\bf (S1)} and (c1),
$$\left|\int_{E_*}^{E_{**}} \, (\partial h)(\sin  M \rho^{(M)}-\sin M \rho) \,  dE\right|\leq 5\varepsilon_0^{\frac\sigma2}\cdot |M| \cdot \varepsilon_J^{\frac14}  \leq \varepsilon_J^{\frac18}.$$
\begin{itemize}
  \item For $(E_*, E_{**})\subset\Gamma_0^{(M)}$, assume $M\neq\pm 1$ and take $M=\pm 1$ as trivial cases. Note that $\rho^{(M)}=\xi^{(M)}$ on $\Gamma_0^{(M)}$. To compute  $\int_{E_*}^{E_{**}} \, (\partial h)\sin  M \rho^{(M)}\,  dE$, we assume that $\sin\xi^{(M)}\neq0$, hence, by {\bf (S2)}, we have
  $\partial\rho^{(M)}=-\frac{\partial {\rm tr}A^{(M)}}{2\sin\rho^{(M)}}$. Then
      \begin{eqnarray*}
       && \int_{E_*}^{E_{**}} (\partial h)\sin  M \rho^{(M)} \,  dE \\
       &=& -2\int_{E_*}^{E_{**}} \frac{\partial h}{\partial{\rm tr} A^{(M)}}\sin  M \rho^{(M)}\,  \sin\rho^{(M)} \cdot\partial\rho^{(M)}  dE\\
       &=& -\int_{E_*}^{E_{**}} \frac{\partial h}{\partial{\rm tr} A^{(M)}}\left[\cos(M-1)\rho^{(M)}-\cos(M+1)\rho^{(M)}\right]\partial\rho^{(M)}  dE\\
       &=& -\left. \frac{\partial h}{\partial{\rm tr}A^{(M)}}\left[\frac{\sin(M-1)\rho^{(M)}}{M-1}-\frac{\sin(M+1)\rho^{(M)}}{M+1}\right]\right|_{(E_*, \,  E_{**})}\\
       &&+ \, \int_{E_*}^{E_{**}} \partial\left(\frac{\partial h}{\partial{\rm tr}A^{(M)}} \right) \left[\frac{\sin(M-1)\rho^{(M)}}{M-1}-\frac{\sin(M+1)\rho^{(M)}}{M+1}\right] dE.
      \end{eqnarray*}
This can be bounded by $\frac{\varepsilon_0^{\frac18}}{|M|}$, since (\ref{sigma_m_0}) implies that
$$|\partial{\rm tr} A^{(M)}+1|_{\Gamma_0^{(M)}}, \; |\partial^2{\rm tr}A^{(M)}|_{\Gamma_0^{(M)}} \leq 2\varepsilon_0^{\frac13}.$$
  \item For $(E_*, E_{**})\subset\Gamma_{j+1}^{(M)}$, by {\bf (S2)}, there is one interval ${\cal I}\subset (E_*, \,  E_{**})$,
  such that
  $\xi^{(M)}=0$. So (c1) implies $\partial h= 0$ on ${\cal I}$. On $(E_*, \,  E_{**})\setminus {\cal I}$, noting that $\partial\rho^{(M)}=\partial\xi^{(M)}$ and in view of (\ref{esti_plat}), we have
$\left| \frac {\partial h}{\partial \rho^{(M)}} \right|\leq 3 \varepsilon_j^{\frac\sigma3}|\sin\xi^{(M)}|^3 \leq \frac{1}{12}\varepsilon_0^{\frac\sigma4}$, and
  $$\left|\partial\left(\frac {\partial h}{\partial \rho^{(M)}}\right)\right|\leq\left|\frac {\partial^2 h}{\partial \rho^{(M)}}\right| +\frac {\left|\partial h\cdot \partial^2 \rho^{(M)}\right|}{|\partial \rho^{(M)}|^2}\leq3\varepsilon_j^{\frac\sigma3}|\sin\xi^{(M)}| + 18N_j^{8\tau}\varepsilon_j^{\frac\sigma3}\leq\frac{1}{30}\varepsilon_0^{\frac\sigma4}.$$
Therefore, with ${\cal I}_1$ and ${\cal I}_2$ denoting the two connected components of $(E_*, \,  E_{**})\setminus {\cal I}$,
\begin{eqnarray*}
\int_{E_*}^{E_{**}} \, (\partial h)\sin  M \rho^{(M)} \,  dE&=&\int_{(E_*, \,  E_{**})\setminus {\cal I}} \, (\partial h)\sin  M \rho^{(M)} \,  dE\\
&=&\int_{(E_*, \,  E_{**})\setminus {\cal I}} \, \frac {\partial h}{\partial \rho^{(M)}}\sin  M \rho^{(M)} \cdot \partial\rho^{(M)}\, dE  \\
&=&\frac{-1}{M}\left[\left.\frac {\partial h}{\partial \rho^{(M)}}\, \cos M\rho^{(M)}\right|_{{\cal I}_1}+ \left.\frac {\partial h}{\partial \rho^{(M)}}\, \cos M\rho^{(M)}\right|_{{\cal I}_2}\right]\\[1mm]
& &+\, \frac{1}{M}\int_{(E_*, \,  E_{**})\setminus {\cal I}} \, \partial\left(\frac {\partial h}{\partial \rho^{(M)}}\right)\, \cos M\rho^{(M)} \,  dE,
\end{eqnarray*}
which can be bounded by $\frac{\varepsilon_0^{\frac\sigma4}}{2|M|}$.
\end{itemize}
So, for each $(E_*, E_{**})\subset \Gamma^{(M)}$, we have $\left| \int_{E_*}^{E_{**}} (\partial h)\sin M \rho\,  dE\right|\leq \frac{\varepsilon_0^{\frac\sigma4}}{2|M|} + \varepsilon_J^{\frac18}\leq  \frac{\varepsilon_0^{\frac\sigma4}}{|M|}$, and then
\begin{equation}\label{esti_complicated}
\left|\frac{1}{M} \sum_{(E_*, E_{**})\subset \Gamma^{(M)}} \int_{E_*}^{E_{**}} (\partial h)\sin M \rho\,  dE\right|\leq \frac{|\ln\varepsilon_0|^{(J+1)^3 d}\,  \varepsilon_0^{\frac\sigma4} }{M^2}\leq\frac{\varepsilon_0^{\frac\sigma6}}{|M|^{\frac32}}.
\end{equation}
Note that in getting (\ref{esti_complicated}), we need to consider two cases about $M$:
\begin{enumerate}
  \item [(\uppercase\expandafter{\romannumeral1})] If $|M|\leq \varepsilon_0^{-\sigma}$, which means $J(M)=0$, then
$|\ln\varepsilon_0|^{(J+1)^3 d}\, \varepsilon_0^{\frac\sigma4}\leq \varepsilon_0^{\frac\sigma6}$.
  \item [(\uppercase\expandafter{\romannumeral2})] If $|M|> \varepsilon_0^{-\sigma}$, which means $J(M)\geq 1$ and $|M|> \varepsilon_{J-1}^{-\sigma}=\varepsilon_{0}^{-\sigma(1+\sigma)^{J-1}}$, then
$$\frac{|\ln\varepsilon_0|^{(J+1)^3 d}\, \varepsilon_0^{\frac\sigma4}}{|M|^2}\leq \frac{|\ln\varepsilon_0|^{(J+1)^3 d}\, \varepsilon_0^{\frac\sigma4}\cdot\varepsilon_{J-1}^{\frac\sigma2}}{|M|^{\frac32}}\leq \frac{\varepsilon_{J}^{\frac\sigma6}}{|M|^{\frac32}}.$$
\end{enumerate}

As a result, by combining (\ref{esti_simple}) and (\ref{esti_complicated}),
$\displaystyle \left|\int_{\inf\sigma(H)}^{\sup\sigma(H)} h\, \cos M \rho   \cdot \partial\rho\, dE\right|\leq \frac{\varepsilon_0^{\frac{\sigma^2}6}}{|M|^{1+\frac\sigma6}}$.
\qed

\begin{Remark}
The initial aim was to bound the integral $\int_{\inf\sigma(H)}^{\sup\sigma(H)} h\, \cos M \rho \, \partial\rho\, dE$ by $\frac{1}{|M|^2}$.
If $h$ is ${\cal C}^2$ on $[\inf\sigma(H), \sup\sigma(H)]$, we can get this estimate by the integration by parts two times since
 $\frac{h}{M}  \sin M \rho$ vanishes at $\inf\sigma(H)$ and $\sup\sigma(H)$.
But here $h$ is just piecewise ${\cal C}^2$ on $\Gamma^{(M)}$ and it is not continuous at the edge points.
We expect the bound $|M|^{-(1+\frac\sigma6)}$ instead. It also guarantees the convergences of the sum over $M\in\Z\setminus\{0\}$, which will be applied in the next subsection.

On each connected component $(E_*, E_{**})\in\Gamma_{J+1}^{(M)}$, where $\sin M\rho$ does not vanish at both edge points,
$\left.h \right|_{(E_*, E_{**})}$ is well estimated by the ${\cal C}^2$ property.
As for the external variation of $h$, i.e., to control $\left.h \right|_{E^-_*}^{E^+_{**}}$(which is necessary in the integration by parts, as shown in (\ref{sum_of_integral})), we need an additional condition {\rm (c2)}.
This is related to (\ref{edge_point}) in {\bf (S4)} of Proposition \ref{propsana1} and the last statement of Lemma \ref{coefficient_abc}.
\end{Remark}

Back to the Bloch-waves and their approximations constructed in Subsection \ref{subsec_bloch}.
From now on till the end of this section, we fix $\theta\in(2\T)^d$ and we shall not report this dependence explicitly.

As a direct application of Lemma \ref{hmn_integral}, we have
\begin{Lemma}\label{lemma_m-n}
For $m,n\in\Z$, $M\in\Z\setminus\{0\}$,
$\displaystyle \left|\int_{\Sigma} \beta_{m, m_\Delta} \beta_{n, n_\Delta}\cos M\rho\cdot\partial\rho\, dE\right|\leq \frac{\varepsilon_0^{\frac{\sigma^2}{7}}}{|M|^{1+\frac\sigma6}}$.
\end{Lemma}
\proof By Lemma \ref{coefficient_abc} and (\ref{error_beta_n}), we get the following properties of $\beta^{(M)}_{m, m_\Delta} \beta^{(M)}_{n, n_\Delta}$ and $\beta_{m, m_\Delta} \beta_{n, n_\Delta}$:
 \begin{itemize}
     \item [(p1)] For $\nu=0,1,2$,
     $\displaystyle \left\{\begin{array}{ll}
                                          |\partial^\nu(\beta^{(M)}_{m, m_\Delta} \beta^{(M)}_{n, n_\Delta} - \delta_{m, m_\Delta}\delta_{n, n_\Delta})|_{\Gamma^{(M)}_0}\leq 3\varepsilon_{0}^{\frac14}, & \\[1mm]
                                          |\partial^{\nu} (\beta^{(M)}_{m, m_\Delta} \beta^{(M)}_{n, n_\Delta})|\leq \varepsilon_j^{\frac\sigma3}|\sin\xi^{(M)}|^{5-2\nu} \;\ {\rm on} \;\  \Gamma^{(M)}_{j+1}, &  0\leq j\leq J
                                        \end{array}
      \right.$,
     \item [(p2)] $\left|\left. \beta^{(M)}_{m, m_\Delta} \beta^{(M)}_{n, n_\Delta}\right|^{E^+_{**}}_{E^-_*}\right|\leq \varepsilon_{J}^{\sigma(1+\frac\sigma6)}$ for any connected component $(E_*, E_{**})\subset \Gamma^{(M)}_{J+1}$,
     \item [(p3)] $|\beta_{m, m_\Delta} \beta_{n, n_\Delta}-\beta^{(M)}_{m, m_\Delta} \beta^{(M)}_{n, n_\Delta}|_{\Sigma_j} \leq \varepsilon_{J}^{\frac15}$, $0\leq j\leq J+1$.
   \end{itemize}
Hence, to compute the integral $\int_{\Sigma} \beta_{m, m_\Delta} \beta_{n, n_\Delta} \cos M\rho\cdot\partial\rho\, dE$, we can consider the integral $\int_{\Sigma} \beta^{(M)}_{m, m_\Delta} \beta^{(M)}_{n, n_\Delta} \cos M\rho\cdot\partial\rho\, dE$ instead.
Indeed, (p3) implies that
$$\sum_{j=0}^{J+1}\left|\int_{\Sigma_j}(\beta_{m, m_\Delta} \beta_{n, n_\Delta}-\beta^{(M)}_{m, m_\Delta}\beta^{(M)}_{n, n_\Delta} ) \cos M \rho   \cdot \partial\rho\, dE\right|\leq \varepsilon_J^{\frac16}.$$
Combining the fact that $|\rho(\Sigma_{j+1})|\leq |\ln\varepsilon_0|^{(j+1)^3 d}\varepsilon_j^{\sigma}\leq \varepsilon_j^{\frac{2\sigma}{3}}$, and recalling that $J=J(M)=\min\{j\in\N: |M|\leq \varepsilon_j^{-\sigma}\}$,
\begin{equation}\label{int_appro}
\left|\int_{\Sigma}(\beta_{m, m_\Delta} \beta_{n, n_\Delta}-\beta^{(M)}_{m, m_\Delta}\beta^{(M)}_{n, n_\Delta} ) \cos M \rho   \cdot \partial\rho\, dE\right| \leq \varepsilon_{J+1}^{\frac{\sigma}{2}} \leq \frac{\varepsilon_J^{\frac{\sigma^2}{4}}}{|M|^{1+\frac\sigma4}}.
\end{equation}

By the gap-labelling theorem mentioned in Subsection \ref{pre_op_cocycle}, $\partial\rho=0$ on $\R\setminus\sigma(H)$.
So we apply Lemma \ref{hmn_integral} to the approximated integral
$$\int_{\Sigma}\beta^{(M)}_{m, m_\Delta} \beta^{(M)}_{n, n_\Delta} \cos M \rho   \cdot \partial\rho\, dE=
\int_{\inf\sigma(H)}^{\sup\sigma(H)}\beta^{(M)}_{m, m_\Delta} \beta^{(M)}_{n, n_\Delta} \cos M \rho   \cdot \partial\rho\, dE,$$
with the conditions (c1) and (c2) verified by (p1) and (p2) respectively,
and get
$$
\left|\int_{\inf\sigma(H)}^{\sup\sigma(H)} \beta^{(M)}_{m, m_\Delta} \beta^{(M)}_{n, n_\Delta} \cos M \rho   \cdot \partial\rho\, dE\right|\leq \frac{\varepsilon_0^{\frac{\sigma^2}{6}}}{|M|^{1+\frac\sigma6}}.
$$
Together with (\ref{int_appro}), the proof is finished.\qed

\subsection{Modified spectral transformation}\label{modified_spec_trans}
\noindent

For Schr\"odinger operator $H$,
we define the modified spectral transformation ${\cal S}$ on $\ell^2(\Z)$:
$${\cal S} q=\left(\begin{array}{c}
                                                         \sum_{n\in\Z}q_n {\Cal K}_n \\[1mm]
                                                         \sum_{n\in\Z}q_n {\Cal J}_n
                                                       \end{array}
     \right).$$
Let the matrix of measures $d\varphi$ be
$$
\left . d\varphi \right|_\Sigma :=\frac1\pi\left(\begin{array}{cc}
                                         (\partial\rho)^{-1} & 0 \\[1mm]
                                         0 & (\partial\rho)^{-1}
                                       \end{array}\right)\, dE, \quad \left. d\varphi\right|_{\R\setminus \Sigma} := 0.
$$
Recall the definition of ${\cal L}^2-$space given in (\ref{gen_L2}).
So here ${\cal L}^2(d\varphi)$ means the space of vectors
$G=(g_j)_{j=1,2}$, with $g_j$ functions of $E\in\R$ satisfying
$$
\|G\|_{{\cal L}^2(d\varphi)}^2:=\frac1\pi\int_\Sigma (|g_1|^2+|g_2|^2)\,  (\partial\rho)^{-1}dE<\infty.
$$

The following lemma shows that ${\cal S}$ is well defined on $\ell^2(\Z)$ to ${\cal L}^2(d\varphi)$.
\begin{Lemma}\label{well-defined}
For any $q\in\ell^2(\Z)\setminus\{0\}$, we have
$0<\left\|{\cal S}q\right\|_{{\cal L}^2(d\varphi)}< 3\| q\|_{\ell^2(\Z)}$.
\end{Lemma}
\proof
Let $d\tilde\varphi:=(\partial\rho)^2 d\varphi$, i.e.,
$$
\left . d\tilde\varphi \right|_\Sigma :=\frac1\pi\left(\begin{array}{cc}
                                         \partial\rho & 0 \\[1mm]
                                         0 & \partial\rho
                                       \end{array}\right)\, dE, \quad \left. d\tilde\varphi\right|_{\R\setminus \Sigma} := 0.
$$
To bound $\|{\cal S}q\|_{{\cal L}^2(d\varphi)}$, we can bound $\|{\cal S}q\|_{{\cal L}^2(d\tilde\varphi)}$ instead.
Indeed, since $(2\sin\rho)^{-1}< \partial\rho<\infty$ on $\Sigma$, we have
$\|{\cal S}q\|_{{\cal L}^2(d\varphi)}\leq 2 \|{\cal S}q\|_{{\cal L}^2(d\tilde\varphi)}$.

Recall that
$\left(\begin{array}{c}
{\cal K}_n \\[1mm]
{\cal J}_n
\end{array}\right)=\left(\begin{array}{c}
\sum_{n_\Delta} \beta_{n, n_\Delta} \sin n_\Delta \rho \\[1mm]
\sum_{n_\Delta} \beta_{n, n_\Delta} \cos n_\Delta \rho
\end{array}\right)$ on $\Sigma$.
Given any $q\in\ell^2(\Z)$ with finite support, saying $[-N, N]$, we have
\begin{eqnarray*}
\|{\cal S}q\|^2_{{\cal L}^2(d\tilde\varphi)} &=&\frac{1}{\pi}\sum_{|m|,|n| \leq N} q_m\bar q_n \int_{\Sigma}( {\Cal K}_m\,  {\Cal K}_n+  {\Cal J}_m\, {\Cal J}_n)\, \partial\rho \,dE\\
&=&\frac{1}{\pi} \sum_{|m|,|n| \leq N} q_m\bar q_n \sum_{m_\Delta,\, n_\Delta} \int_{\Sigma} \beta_{m, m_\Delta}\beta_{n, n_\Delta} \cos(m_\Delta-n_\Delta)\rho\cdot  \partial\rho \, dE.
\end{eqnarray*}
Applying Lemma \ref{lemma_nn} and \ref{lemma_m-n} to the above integral, corresponding to the cases $m_\Delta-n_\Delta=0$ and $m_\Delta-n_\Delta\neq0$ respectively, we have
$$\left\{ \begin{array}{ll}
           \displaystyle \left|\frac1\pi\int_{\Sigma}({\Cal K}^2_n+ {\Cal J}^2_n) \,  \partial\rho \, dE-1\right|\leq \varepsilon_0^{\frac{\sigma^2}{8}},  &  \\[1mm]
           \displaystyle\left|\frac1\pi\int_{\Sigma}({\Cal K}_m\, {\Cal K}_n+ {\Cal J}_m\, {\Cal J}_n) \,  \partial\rho \, dE\right|\leq \frac{\varepsilon_0^{\frac{\sigma^2}{8}}}{|m-n|^{1+\frac\sigma6}},  & m\neq n
          \end{array}
 \right. .$$
Then we can get
\begin{eqnarray}
\left|\frac1\pi\sum_{|n|\leq N}|q_n|^2\int_{\Sigma}({\Cal K}_n^2+{\Cal J}_n^2)\,  \partial\rho \, dE- \sum_{|n|\leq N}|q_n|^2\right|&\leq& \varepsilon_0^{\frac{\sigma^2}{8}}\sum_{|n|\leq N}|q_n|^2,\label{l2sum_1}\\
\left|\frac1\pi\sum_{|m|,|n| \leq N\atop{m\neq n}}q_m\bar q_n\int_{\Sigma} ({\Cal K}_m\, {\Cal K}_n+{\Cal J}_m\, {\Cal J}_n) \,  \partial\rho \, dE \right|
&\leq&  \varepsilon_0^{\frac{\sigma^2}{8}}\sum_{k\in\Z\setminus\{0\}} \frac{1}{|k|^{1+\frac\sigma6}}\sum_{n\in\Z}|q_{n+k}| | \bar q_n| \nonumber\\
&\leq&  \varepsilon_0^{\frac{\sigma^2}{9}} \sum_{n\in\Z}|q_n|^2. \label{l2sum_3}
\end{eqnarray}
Note that to get (\ref{l2sum_3}), we have applied H\"older's inequality for each given $k$:
$$\sum_{n\in\Z}|q_{n+k}| | \bar q_n|\leq \left(\sum_{n\in\Z}|q_{n+k}|^2\right)^{\frac12} \left(\sum_{n\in\Z}|q_{n}|^2\right)^{\frac12}=\sum_{n\in\Z}|q_n|^2.$$
By combining (\ref{l2sum_1}) and (\ref{l2sum_3}), we have
\begin{equation}\label{two_bounds}
\left(1-\varepsilon_0^{\frac{\sigma^2}{10}}\right)\|q\|^2_{\ell^2(\Z)}\leq \|{\cal S}q\|^2_{{\cal L}^2(d\tilde\varphi)}\leq\left(1+\varepsilon_0^{\frac{\sigma^2}{10}}\right)\|q\|^2_{\ell^2(\Z)}.
\end{equation}
Since any $q\in\ell^2(\Z)$ can be approximated by finitely supported vectors in the sense of $\ell^2$, we can pass (\ref{two_bounds}) to any $q\in\ell^2(\Z)$.
Hence, $\|{\cal S}q\|_{{\cal L}^2(d\varphi)}< 2\|{\cal S}q\|_{{\cal L}^2(d\tilde\varphi)}< 3\|q\|_{\ell^2(\Z)}$.

Note that the measure $(\partial \rho)^{-1}dE$ is absolutely continuous with respect to $\partial \rho \, dE$ and $\partial \rho$ is positive everywhere on $\Sigma$. By (\ref{two_bounds}), we have that $\|{\cal S}q\|^2_{{\cal L}^2(d\varphi)}>0$ if $q\neq 0$. \qed

We can see that ${\cal K}_n$ and ${\cal J}_n$ are differentiable in the sense of Whitney on each $\Sigma_j$ and
\begin{equation}\label{derivative_KnJn}
\left(\begin{array}{c}
\partial{\cal K}_n \\[1mm]
\partial{\cal J}_n
\end{array}\right)=\left(\begin{array}{c}
\hat {\cal K}_n \\[1mm]
\hat {\cal J}_n
\end{array}\right)+\left(\begin{array}{c}
\sum_{n_\Delta} (\partial  \beta_{n, n_\Delta}) \sin n_\Delta \rho \\[1mm]
\sum_{n_\Delta} (\partial  \beta_{n, n_\Delta}) \cos n_\Delta \rho
\end{array}\right),
\end{equation}
where $\partial  \beta_{n, n_\Delta}$ is the derivative in the sense of Whitney on $\Sigma_j$, and
$$\left.\left(\begin{array}{c}
\hat {\cal K}_n \\[1mm]
\hat {\cal J}_n
\end{array}\right)\right|_{\Sigma}
:=\left(\begin{array}{c}
\partial\rho\sum_{n_\Delta}  n_\Delta \beta_{n, n_\Delta} \cos n_\Delta \rho \\[1mm]
-\partial\rho \sum_{n_\Delta}  n_\Delta \beta_{n, n_\Delta} \sin n_\Delta \rho
\end{array}\right).
$$
Since $\{\Sigma_j\}_{j\geq0}$ are mutually disjoint, $\partial  \beta_{n, n_\Delta}$ and hence $\partial{\cal K}_n$, $\partial{\cal J}_n$ are well defined on $\Sigma$.

\begin{Remark}
As shown in Subsection \ref{pre_op_cocycle},
the classical spectral transformation is a unitary transformation from $\ell^2(\Z)$ to ${\cal L}^2(d\mu)$, with $d\mu$ the matrix of spectral measures introduced by $m-$functions.
In contrast, to get better differentiability with respect to $E$, the modified spectral transformation ${\cal S}$ here is not a unitary one.
Comparing with (\ref{classical_uv}) for the free Schr\"odinger operator, ${\cal K}_n$ and ${\cal J}_n$ for ${\cal S}$ have no divisor as `` $\sim\sin\rho$" and they have a smoothing factor $\sin^{10}\xi$ in a small part of spectrum to cover the singularities.
Moreover, instead of the spectral measures shown in Theorem \ref{spectral_measure_matrix},
we use the explicit measure $(\partial \rho)^{-1}dE$, which has a nice regularity in view of the transversality (\ref{transversality_rho}) of $\partial \rho$.
\end{Remark}

\begin{Remark}
With the purely absolute continuity of the spectrum, we can conclude that the spectral transformation for any non-zero $q\in\ell^2(\Z)$ is supported on a subset of $\sigma(H)$ with positive Lebesgue measure.
Hence, in constructing the modified spectral transformation, we can neglect a zero-measure subset of $\sigma(H)$ and just focus on $\Sigma$. This is the necessity of the purely absolute continuity in the proof.
\end{Remark}

\begin{Lemma}\label{lemma_sobolevnorm}
For any $q\in\ell^2(\Z)$ with $\|q\|_D<\infty$,
  \begin{equation}\label{H1_norm-whole}
\left|\left\|\left(\begin{array}{c}
                   \sum_{n} q_n\partial{\cal K}_n \\[1mm]
                    \sum_{n} q_n\partial{\cal J}_n
                 \end{array}
\right)\right\|_{{\cal L}^2(d\varphi)} - \|q\|_D\right|
\leq \varepsilon_0^{\frac\sigma4}|q_0|+ \varepsilon_0^{\frac{\sigma^2}{10}} \|q\|_D.
\end{equation}
\end{Lemma}
\proof We decompose $\left(\begin{array}{c}
                   \sum_{n} q_n\partial{\cal K}_n \\[1mm]
                    \sum_{n} q_n\partial{\cal J}_n
                 \end{array}
\right)$ into
$$\left(\begin{array}{c}
                    q_0\, \partial{\cal K}_0 \\[1mm]
                    q_0\, \partial{\cal J}_0
                 \end{array}
\right)+\left(\begin{array}{c}
                   \sum_{n\in\Z\setminus\{0\}} q_n(\partial{\cal K}_n-\hat{\cal K}_n) \\[1mm]
                    \sum_{n\in\Z\setminus\{0\}} q_n(\partial{\cal J}_n-\hat{\cal J}_n)
                 \end{array}
\right)+\left(\begin{array}{c}
                   \sum_{n\in\Z\setminus\{0\}} q_n \hat{\cal K}_n \\[1mm]
                    \sum_{n\in\Z\setminus\{0\}} q_n \hat{\cal J}_n
                 \end{array}
\right).$$

By a direct computation, we can see, from (\ref{K0J0}) and (\ref{esti_abc_sigma_avant}), that
\begin{equation}\label{q_0}
\left\|\left(\begin{array}{c}
                    q_0\, \partial{\cal K}_0 \\[1mm]
                    q_0\, \partial{\cal J}_0
                 \end{array}
\right)\right\|_{{\cal L}^2(d\varphi)}^2 = \left\|\left(\begin{array}{c}
                    0 \\[1mm]
                    q_0(\partial\beta_{0,0}+2(\partial\beta_{0,1})\cos\rho-2(\partial\rho)\beta_{0,1}\sin\rho)
                 \end{array}
\right)\right\|_{{\cal L}^2(d\varphi)}^2\leq\varepsilon_0^{\frac\sigma2}|q_0|^2.
\end{equation}
In view of (\ref{esti_abc_sigma_avant}) and (\ref{derivative_KnJn}), we see $|\hat {\Cal K}_n-\partial  {\Cal K}_n|_{\Sigma_j}$, $|\hat {\Cal J}_n-\partial  {\Cal J}_n|_{\Sigma_j}\leq \varepsilon_0^{\frac\sigma2}$ for every $\Sigma_j$,
 so
\begin{eqnarray}
 &&\left\|\left(\begin{array}{c}
                   \sum_{n\neq0} q_n(\hat {\Cal K}_n-\partial  {\Cal K}_n) \\[1mm]
                    \sum_{n\neq0} q_n(\hat {\Cal J}_n-\partial  {\Cal J}_n)
                 \end{array}
\right)\right\|_{{\cal L}^2(d\varphi)}^2 \nonumber\\
 &\leq & \sum_{m, n\in\Z\setminus\{0\}\atop{j\geq 0}}\int_{\Sigma_j}|q_m| |\bar q_n|\left( \left|\hat {\Cal K}_m-\partial  {\Cal K}_m\right| \left|\hat {\Cal K}_n-\partial  {\Cal K}_n \right| +\left|\hat {\Cal J}_m-\partial  {\Cal J}_m\right| \left|\hat {\Cal J}_n-\partial  {\Cal J}_n\right|\right) \frac{(\partial\rho)^{-1}}{\pi}dE   \nonumber\\
&\leq& \frac{2\varepsilon_0^{\sigma}}{\pi}\sum_{m, n\in\Z\setminus\{0\}}|q_m| |\bar q_n| \int_{\Sigma} (\partial\rho)^{-1} dE \nonumber\\
&\leq& \varepsilon_0^{\frac\sigma2}\sum_{n\in\Z\setminus\{0\}} n^2|q_n |^2.  \label{error}
\end{eqnarray}

To consider the third part, we first assume that $q$ has finite support, saying $[-N,N]$.
So we have
$$\left\|\left(\begin{array}{c}
                   \sum_{0<|n|\leq N} q_n \hat {\Cal K}_n  \\[1mm]
                   \sum_{0<|n|\leq N} q_n \hat {\Cal J}_n
                 \end{array}
\right)\right\|_{{\cal L}^2(d\varphi)}^2= \frac1\pi\sum_{{0<|m|, |n|\leq N}} mnq_m \bar q_n \int_\Sigma\frac{\hat {\Cal K}_m \hat {\Cal K}_n+\hat {\Cal J}_m \hat {\Cal J}_n}{mn}(\partial\rho)^{-1} dE.$$
In view of the definition of $\hat {\Cal K}_n$, $\hat {\Cal J}_n$, we have, for $m,n\in\Z\setminus\{0\}$,
$$\int_{\Sigma}\frac{\hat {\Cal K}_m \hat {\Cal K}_n+\hat {\Cal J}_m \hat {\Cal J}_n}{mn}(\partial\rho)^{-1} dE=\sum_{m_\Delta,\, n_\Delta}\frac{m_\Delta n_\Delta}{mn}\int_\Sigma \beta_{m, m_\Delta}\beta_{n, n_\Delta} \cos(m_\Delta-n_\Delta)\rho\cdot  \partial\rho \, dE.$$
Applying Lemma \ref{lemma_nn} and \ref{lemma_m-n} to the above integral, corresponding to the cases $m_\Delta-n_\Delta=0$ and $m_\Delta-n_\Delta\neq0$ respectively, and noting that $\left|\frac{m_\Delta n_\Delta}{mn}\right|\leq 2$ for any $m,n\neq0$, we get
$$\left\{ \begin{array}{ll}
           \displaystyle \left|\frac1\pi\int_{\Sigma}\frac{\hat {\Cal K}^2_n+ \hat {\Cal J}^2_n}{n^2} (\partial\rho)^{-1} dE-1\right|\leq \varepsilon_0^{\frac{\sigma^2}{8}},  &  \\[1mm]
           \displaystyle\left|\frac1\pi\int_{\Sigma}\frac{\hat {\Cal K}_m\, \hat {\Cal K}_n+\hat {\Cal J}_m\, \hat {\Cal J}_n}{mn} (\partial\rho)^{-1} dE\right|\leq \frac{\varepsilon_0^{\frac{\sigma^2}{8}}}{|m-n|^{1+\frac\sigma6}},  & m\neq n
          \end{array}
 \right. .$$
Therefore, similar to (\ref{l2sum_1}) and (\ref{l2sum_3}), we have
$$\left|\frac1\pi\sum_{0<|n|\leq N}n^2|q_n|^2\int_{\Sigma}\frac{\hat {\Cal K}^2_n+ \hat {\Cal J}^2_n}{n^2} (\partial\rho)^{-1}dE- \sum_{|n|\leq N}n^2|q_n|^2\right|\leq \varepsilon_0^{\frac{\sigma^2}{8}}\sum_{|n|\leq N}n^2|q_n|^2,$$
\begin{eqnarray*}
   & & \left|\frac1\pi\sum_{0<|m|, |n|\leq N\atop{m\neq n}} mn q_m\bar q_n\int_{\Sigma} \frac{\hat {\Cal K}_m\, \hat {\Cal K}_n+\hat {\Cal J}_m\, \hat {\Cal J}_n}{mn} (\partial\rho)^{-1}dE \right| \\
 &\leq &  \varepsilon_0^{\frac{\sigma^2}{8}}\sum_{k\in\Z\setminus\{0\}} \frac{1}{|k|^{1+\frac\sigma6}}\sum_{n\in\Z}|(n+k)q_{n+k}|
|n \bar q_n|\\
&\leq& \varepsilon_0^{\frac{\sigma^2}{9}} \sum_{n\in\Z}n^2|q_n|^2.
\end{eqnarray*}
Because of these two inequalities, we get
\begin{equation}\label{two_bounds_Diff}
\left(1-\varepsilon_0^{\frac{\sigma^2}{10}}\right)\|q\|^2_D\leq\left\|\left(\begin{array}{c}
                   \sum_{n\in\Z\setminus\{0\}} q_n \hat {\Cal K}_n  \\[1mm]
                   \sum_{n\in\Z\setminus\{0\}} q_n \hat {\Cal J}_n
                 \end{array}
\right)\right\|^2_{{\cal L}^2(d\varphi)}\leq \left(1+\varepsilon_0^{\frac{\sigma^2}{10}}\right)\|q\|^2_D.
\end{equation}
Since any $q\in\ell^2(\Z)$ with $\|q\|_D<\infty$ can be approximated by finitely supported vectors in the sense of $\|\cdot\|_D$, we can pass the inequality (\ref{two_bounds_Diff}) to any $q\in\ell^2(\Z)$ with $\|q\|_D<\infty$.

Combining (\ref{q_0}), (\ref{error}) and (\ref{two_bounds_Diff}), we get (\ref{H1_norm-whole}).\qed


The following lemma shows that $\left(\begin{array}{c}
                   \sum_{n} q_n\partial{\cal K}_n \\[1mm]
                    \sum_{n} q_n\partial{\cal J}_n
                 \end{array}
\right)$ converges to the derivative of the modified spectral transformation under some suitable condition.

\begin{Lemma}\label{orderchanging}
For $q\in\ell^2(\Z)$ satisfying $\|q\|_D<\infty$, with
\begin{itemize}
  \item [(a1)]
  $\left(\begin{array}{c}
                   \sum_{n\in\Z} q_n {\cal K}_n \\[1mm]
                    \sum_{n\in\Z} q_n {\cal J}_n
                 \end{array}
\right)$ convergent to $F=\left(\begin{array}{c}
                   F_1 \\[1mm]
                   F_2
                 \end{array}\right)$ uniformly in $E$,
  \item [(a2)] $\left(\begin{array}{c}
                   \sum_{n\in\Z} q_n\partial{\cal K}_n \\[1mm]
                    \sum_{n\in\Z} q_n\partial{\cal J}_n
                 \end{array}
\right)$ convergent to $\tilde H=\left(\begin{array}{c}
                   \tilde H_1 \\[1mm]
                   \tilde H_2
                 \end{array}\right)$ in the sense of ${\cal L}^2( d\varphi)$,
\end{itemize}
if $F$ is ${\cal C}^1_W$ on each $\Sigma_j$, then $\partial F= \tilde H$ a.e. on $\Sigma$.
\end{Lemma}
\proof Let $\beta_{n,n_\Delta}^j$ be the extension of $\beta_{n,n_\Delta}$, ${\cal C}^1$ on $[\inf\sigma(H), \sup\sigma(H)]$, with $\beta_{n,n_\Delta}^j|_{\Sigma_j}=\beta_{n,n_\Delta}$, and let ${\cal K}^j_n:=\sum_{n_\Delta}\beta_{n,n_\Delta}^j\sin n_\Delta\rho$, ${\cal J}^j_n:=\sum_{n_\Delta}\beta_{n,n_\Delta}^j\cos n_\Delta\rho$.

Obviously, ${\cal K}^j_n$ is absolutely continuous on $[\inf\sigma(H), \sup\sigma(H)]$, so for any ${\cal C}^1$ function $\phi$ on $[\inf\sigma(H), \sup\sigma(H)]$, by the integration by parts,
$$\int_{\Sigma_j} \partial {\Cal K}_n \cdot \phi\, dE =\left. {\Cal K}_n \cdot \phi \right|_{\Sigma_j}  - \int_{\Sigma_j} {\Cal K}_n \cdot \partial \phi\, dE.$$
Here $\Sigma_j$ is a Borel set contained in $\sigma(H)$. It can be written as $$\Sigma_j=[\inf\sigma(H), \sup\sigma(H)]\setminus \bigcup_{l\geq 0} I_{l},$$ with $\{I_{l}\}_{l\geq 0}$ a sequence of intervals, mutually disjoint, and $\left. {\Cal K}_n \cdot \phi \right|_{\Sigma_j}$ is interpreted as
$$\left. {\Cal K}_n \cdot \phi \right|_{\Sigma_j}=\left. {\Cal K}^j_n \cdot \phi \right|_{[\inf\sigma(H), \,  \sup\sigma(H)]}-\sum_{l\geq 0}\left. {\Cal K}^j_n \cdot \phi \right|_{I_l}.$$
Since $\beta^j_{n,n_\Delta}$, $\phi$ and $\rho$ are all absolutely continuous on $[\inf\sigma(H), \,  \sup\sigma(H)]$, we can see the absolute convergence of $\sum_{l\geq 0}\left. {\Cal K}^j_n \cdot \phi \right|_{I_l}$. Hence, by Fubini's theorem,
$$ \sum_n q_n (\left.{\Cal K}_n\cdot  \phi \right|_{\Sigma_j})=\left.\left(\sum_n q_n{\Cal K}_n\right)\cdot  \phi \right|_{\Sigma_j}= \left. F_1 \cdot \phi \right|_{\Sigma_j}.$$
On the other hand, for each $\Sigma_j$, we have, by (a2),
$$\int_{\Sigma_j} |\sum_{|n|\leq N} q_n \partial{\Cal K}_n  - \tilde H_1 |\cdot |\phi| dE
\leq\left(\int_{\Sigma_j} |\sum_{|n|\leq N} q_n \partial{\Cal K}_n - \tilde H_1|^2 (\partial\rho)^{-1} dE\right)^{\frac12} \left(\int_{\Sigma_j} |\phi|^2\partial\rho\, dE\right)^{\frac12},$$
which goes to $0$ as $N\rightarrow\infty$.
Hence,
 $$\int_{\Sigma_j} \partial F_1\cdot\phi\, dE= \left. F_1 \cdot \phi \right|_{\Sigma_j}  - \int_{\Sigma_j} F_1 \cdot \partial \phi\, dE= \sum_n q_n (\left.{\Cal K}_n\cdot  \phi \right|_{\Sigma_j})-  \sum_n q_n\int_{\Sigma_j} {\Cal K}_n \cdot \partial \phi\, dE,$$
 which equals to
 $\sum_n q_n \int_{\Sigma_j} \partial {\Cal K}_n \cdot \phi\, dE=\lim_{N\rightarrow \infty} \int_{\Sigma_j} \sum_{|n|\leq N} q_n \partial {\Cal K}_n \cdot \phi\, dE=\int_{\Sigma_j} \tilde H_1 \cdot \phi\, dE$.
So $\partial F_1=\tilde H_1$ a.e. on each $\Sigma_j$, hence a.e. on $\Sigma$. Similarly, $\partial F_2=\tilde H_2$ a.e. on $\Sigma$. \qed

%

\subsection{Proof of Theorem \ref{thm-qp}}\label{proof_of_thm}
\noindent

Now, let $q(t)=(q_n(t))_{n\in\Z}$ be the solution to the dynamical equation ${\rm i}\dot{q}=H q$, with $q(0)\in \ell^2(\Z)$. Let $G(E,t):=({\cal S}q)(E,t)$. Since, for any $E\in\Sigma$,
$$\frac{1}{\delta}(G(E,t+\delta)-G(E,t))=\frac{1}{\delta}\left(\begin{array}{c}
\sum_n [q_n(t+\delta)-q_n(t)] {\cal K}_n(E) \\[1mm]
\sum_n [q_n(t+\delta)-q_n(t)] {\cal J}_n(E)
\end{array}\right) \ {\rm for} \ \delta >0,$$
we can verify the differentiability of $G(E,t)$ with respect to $t$.
For $E\in\Sigma$,
$${\rm i}\partial_t G(E,t)=\left(\begin{array}{c}
\sum_n (Hq)_n(t) {\cal K}_n(E) \\[1mm]
\sum_n (Hq)_n(t) {\cal J}_n(E)
\end{array}\right)=\left(\begin{array}{c}
\sum_n q_n(t) (H{\cal K})_n(E)\\[1mm]
\sum_n q_n(t) (H{\cal J})_n(E)
\end{array}\right)=E G(E,t), $$
so $G(E,t)= e^{-{\rm i}Et} G(E,0)$.

\begin{Corollary}\label{sum_derivation_exchange}
For any solution $q(t)=(q_n(t))_{n\in\Z}$ to the equation ${\rm i}\dot{q}=H q$, with $q(0)=(q_n(0))_{n\in\Z}$ supported on a finite subset $\Lambda\subset \Z$, we have, for a.e. $E\in\Sigma$,
  \begin{equation}\label{verif_changing_whole}
  \left(\begin{array}{c}
                   \sum_{n\in\Z} q_n(t)\partial{\cal K}_n(E) \\[1mm]
                    \sum_{n\in\Z} q_n(t)\partial{\cal J}_n(E)
                 \end{array}
\right) = -{\rm i} t\cdot e^{-{\rm i}Et} G(E,0) + e^{-{\rm i}Et}\partial G(E,0).
  \end{equation}
\end{Corollary}
\proof $q(0)$ is finitely supported, so $\partial G(E,0)$ is well defined on each $\Sigma_j$, with
$$\partial G(E,0)=\left(\begin{array}{c}
                   \sum_{n\in\Lambda} q_n(0)\partial{\cal K}_n(E) \\[1mm]
                    \sum_{n\in\Lambda} q_n(0)\partial{\cal J}_n(E)
                 \end{array}
\right).$$
Hence, $G(E,t)= e^{-{\rm i}Et} G(E,0)$ is differentiable in the sense of Whitney on each $\Sigma_j$, with
$$\partial G(E,t)=-{\rm i} t\cdot e^{-{\rm i}Et} G(E,0) + e^{-{\rm i}Et}\partial G(E,0).$$
For any finite $t$, $\sum_{n} n^2 |q_n(t)|^2<\infty$, which implies $\sum_{n} |q_n(t)|<\infty$. The $\ell^\infty$ property of ${\cal K}_n$ and ${\cal J}_n$ implies
$$\sum_{n\in\Z} |q_n(t) {\cal K}_n|, \; \sum_{n\in\Z} |q_n(t) {\cal J}_n|\leq 2\sum_{n\in\Z} |q_n(t)|, $$
and by Lemma \ref{lemma_sobolevnorm}, for $N>0$ sufficiently large,
$$\left\|\left(\begin{array}{c}
                   \sum_{|n|>N} q_n(t)\partial{\cal K}_n \\[1mm]
                    \sum_{|n|>N} q_n(t)\partial{\cal J}_n
                 \end{array}
\right)\right\|_{{\cal L}^2(d\varphi)}\leq 2\left(\sum_{|n|>N}n^2|q_n(t)|^2\right)^{\frac12}.$$
So the assumptions (a1) and (a2) of Lemma \ref{orderchanging} are verified.
Applying Lemma \ref{orderchanging}, the proof of (\ref{verif_changing_whole}) is finished.\qed

Given any solution $q(t)$ to ${\rm i}\dot{q}=H q$ with initial datum $q(0)$ satisfying $\|q(0)\|_D<\infty$ and $N>0$, we define $q^N(0)=(q_n^N(0))_n$ as the truncation of $q(0)$, i.e.,
$$q_n^N(0):=
\left\{\begin{array}{cl}
                   q_n(0), & |n|\leq N \\[1mm]
                   0, & |n|>N
                 \end{array}
\right..
$$
By Lemma \ref{well-defined}, $\lim_{N\to\infty}\|{\cal S} q^N(0)\|_{{\cal L}^2(d\varphi)}=\|{\cal S} q(0)\|_{{\cal L}^2(d\varphi)}$.
Let $q^N(t)$ be the solution satisfying ${\rm i}\dot{q}^N=H q^N$ with initial datum $q^N(0)$,
and $G_N(E,t)=({\cal S}q^N)(E,t)$.

In view of Lemma \ref{lemma_sobolevnorm} and Corollary \ref{sum_derivation_exchange}, we can see,
$$
\left|t\|G_N(E,0) \|_{{\cal L}^2(d\varphi)}- \|q^N(t)\|_D\right|
\leq \varepsilon_0^{\frac\sigma4}|q^N_0(t)|+ \varepsilon_0^{\frac{\sigma^2}{10}} \|q^N(t)\|_D+ \|\partial G_N(E,0) \|_{{\cal L}^2(d\varphi)}.
$$
Hence, we have
$$\frac{\|G_N(E,0)\|_{{\cal L}^2(d\varphi)}-t^{-1}{\cal G}_N(t)}{1+\varepsilon_0^{\frac{\sigma^2}{10}}}\leq t^{-1}\|q^N(t)\|_D\leq \frac{\|G_N(E,0)\|_{{\cal L}^2(d\varphi)}+t^{-1}{\cal G}_N(t)}{1-\varepsilon_0^{\frac{\sigma^2}{10}}}$$
with ${\cal G}_N(t):=\|\partial G_N(E,0) \|_{{\cal L}^2(d\varphi)}+\varepsilon_0^{\frac\sigma4}|q^N_0(t)|$.
By Lemma \ref{lemma_sobolevnorm}, Corollary \ref{sum_derivation_exchange}, and the $\ell^2-$conservation law, we can see
$${\cal G}_N(t)\leq 2\left(\sum_{|n|\leq N} n^2|q_n(0)|^2\right)^{\frac12}+\varepsilon_0^{\frac\sigma4}|q^N_0(t)|\leq 2\|q(0)\|_{D} + \varepsilon_0^{\frac\sigma4}\|q(0)\|_{\ell^2(\Z)}.$$
So, for $t$ large enough(independent of $N$), $t^{-1}{\cal G}_N(t)$ goes to zero, and
\begin{equation}\label{growth_finite-supp_whole}
\frac{\|{\cal S} q^N(0)\|_{{\cal L}^2(d\varphi)}}{1+\varepsilon_0^{\frac{\sigma^2}{16}}}\leq t^{-1}\|q^N(t)\|_D \leq \frac{\|{\cal S}q^N(0)\|_{{\cal L}^2(d\varphi)}}{1-\varepsilon_0^{\frac{\sigma^2}{16}}}.
\end{equation}

By the ballistic upper bound (\ref{ballistic_upper_bound}), we have
$$\lim_{N\to\infty}t^{-1}\|q^N(t)-q(t)\|_D\leq 2\lim_{N\to\infty} \|q^N(0)-q(0)\|_{\ell^2(\Z)}+t^{-1}\lim_{N\to\infty} \|q^N(0)-q(0)\|_{D}=0.$$
Combining with the fact that $\lim_{N\to\infty}\|{\cal S} q^N(0)\|_{{\cal L}^2(d\varphi)}=\|{\cal S} q(0)\|_{{\cal L}^2(d\varphi)}$, we can pass
(\ref{growth_finite-supp_whole}) to $N\to\infty$.
Then Theorem \ref{thm-qp} can be proven with
$$C=\|{\cal S} q(0)\|_{{\cal L}^2(d\varphi)}, \quad \zeta=\frac{\sigma^2}{16}.$$

\appendix

\section{Appendix}

\subsection{Proof of Proposition \ref{propsana1}}\label{proofP2}
\noindent

Given $M\in\Z\setminus\{0\}$ with $J=J(M)=\min\left\{j\in\N:|M|\leq \varepsilon_j^{-\sigma}\right\}$.
Recall the iteration process given in the proof of Proposition \ref{propsana} (1) and (2).
To prove Proposition \ref{propsana1}, we just focus on the first $J+1$ steps of iteration.

Assume that $J\geq1$. At the $(j+1)^{\rm th}-$step, $0\leq j \leq J-1$,
as shown in (\ref{H_kj_Aj}), we need a renormalization $H_{k_j, \tilde A_j}$ on the intervals ${\cal I}_{\la k_j\ra}$ where the resonance condition (\ref{resonance_condition_j}) holds.
Then we can construct $\hat Z_{j+1}$ which is close to $H_{k_j, \tilde A_j}$ as in (\ref{appro_hatZ_tildeA_j+1}).
Note that ${\cal I}_{\la k_j\ra}$ is not uniquely determined since we could modify the coefficients in the resonance condition (\ref{resonance_condition_j}) as we need.
So, for the given $M\in\Z\setminus\{0\}$, we can define ${\cal I}_{\la k_j\ra}$ as
$${\cal I}_{\la k_j\ra}:=\left\{E\in\R :- \frac{c_1 \varepsilon_{j}^{\sigma}}{|k_j|^{\tau}} < \xi_{j}-\la k_j\ra<\frac{c_2 \varepsilon_{j}^{\sigma}}{|k_j|^{\tau}}\right\},$$
where $c_1$, $c_2\in [\frac12, 1]$ are two constants, depending on $M$ and $k_j$, such that
$M \rho\left(\partial{\cal I}_{\la k_j\ra}\right)\subset \pi\Z$.
By (\ref{size_I_k_j}), $\rho({\cal I}_{\la k_j\ra})$ is adjustable because $\varepsilon_J^\sigma<\frac{1}{|M|}\leq \varepsilon_{J-1}^\sigma$.

At the $(J+1)^{\rm th}-$step, we can construct $\hat Z_{J+1}$(hence $\tilde Z_{J+1}$ and $\tilde A_{J+1}$) as above,
 with the resonance condition(hence the definition of ${\cal I}_{\la k_J\ra}$) replaced by: \footnote{If $J=0$, we start with the resonance condition (\ref{resonance_J}) directly.}
 \begin{equation}\label{resonance_J}
 {\rm there \  is \ a \  vector} \ 0<|k_J|\leq N_J \ {\rm satisfying } \ \left|\xi_{J}-\la k_J\ra\right|<\frac34\frac{\varepsilon_{J}^{\sigma(1+\frac\sigma2)}}{|k_J|^{\tau}}.
 \end{equation}
Then, as shown above, $|\rho({\cal I}_{\la k_J\ra})|< 2\varepsilon_{J}^{\sigma(1+\frac\sigma2)}$, and,
the slight change in the resonance condition(the index $\sigma$ to $\sigma(1+\frac\sigma2)$), does not affect the estimation as in (\ref{tilde_ZA_j+1}).
Noting that $\partial\xi_{J}>\frac13$, we have $\varepsilon_{J}^{3\sigma(1+\sigma)}\leq|{\cal I}_{\la k_J\ra}|\leq\varepsilon_{J}^{\sigma(1+\frac\sigma3)}$.

Define the sets $\Gamma^{(M)}_j$, $0\leq j \leq J+1$, as
\begin{equation}\label{Gamma^M}
\Gamma^{(M)}_j:=\left\{
\begin{array}{ll}
  \bigcup_{0<|k_J|\leq N_J} {\cal I}_{\la k_J\ra}, & j=J+1 \\[3mm]
  \bigcup_{0<|k_{j-1}|\leq N_{j-1}} {\cal I}_{\la k_{j-1}\ra}\setminus \left(\bigcup_{l=j+1}^{J+1} \Gamma^{(M)}_l\right), & J\geq j \geq 1 \\[3mm]
  \left[\inf\sigma(H), \sup\sigma(H)\right]\setminus \left(\bigcup_{l=1}^{J+1} \Gamma^{(M)}_l\right) ,& j=0
\end{array} \right..
\end{equation}We can get (\ref{measure}) by noting that
$|\rho(\Gamma^{(M)}_{j+1})|\leq \frac1{10} |\ln\varepsilon_0|^{(j+1)^3 d} \cdot 10\varepsilon_j^\sigma$, $0\leq j\leq J$.

Let $\tilde Z^{(M)}:=\tilde Z_{J+1}$
 and $\tilde A^{(M)}:=\tilde A_{J+1}$ which has two eigenvalues $e^{\pm{\rm i}\tilde\alpha^{(M)}}$. Then $\xi^{(M)}:=\Re\tilde\alpha^{(M)}$ is exactly $\xi_{J+1}$.
The finite sequence $\{k^{(M)}_l\}_{0\leq l\leq J}\subset \Z^d$ in {\bf (S1)} is exactly the $k_j$'s given as above, which is piecewise constant, satisfying
$k^{(M)}_{l}=0$ on $\Gamma^{(M)}_{j}$, if $l\geq j$, and $|\xi^{(M)}+ \sum_{l=0}^{J}\la k^{(M)}_l\ra-\rho|\leq \varepsilon_J^{\frac14}$.
\begin{itemize}
  \item On $\Gamma^{(M)}_0$, there is no resonance in these $J+1$ steps, which means $k_j^{(M)}=0$ for every $0\leq j\leq J$,
then each transformation $\hat Z_{j+1}$ is close to identity. So
\begin{equation}\label{sigma_m_0_tilde}
|\partial^\nu (\tilde A^{(M)}- A_0)|\leq \varepsilon_0^{\frac23}, \;\ |\partial^\nu (\tilde Z^{(M)}-Id.)|_{(2\T)^d} \leq \varepsilon_0^{\frac12}, \quad \nu=0,1,2.
\end{equation}
If $\sin\xi^{(M)}\neq 0$, we have ${\rm tr}\tilde A^{(M)}=2\cos\xi^{(M)}$,  so $\partial\xi^{(M)}=-\frac{\partial {\rm tr}\tilde A^{(M)}}{2\sin\xi^{(M)}}$. Similar to the case of Corollary 6 in \cite{E92},
$\partial\xi^{(M)}> \frac13$.
  \item On $\Gamma^{(M)}_{j+1}$, $0\leq j\leq J$, the resonance and the renormalization occur exactly at the $j^{\rm th}-$step, but do not occur afterwards, so for $j+1\leq l\leq J$ and $\nu=0,1,2$, $|\partial^\nu(\hat Z_{l+1}-Id.)|_{(2\T)^d}< \varepsilon_l^{\frac12}$, $|\partial^{\nu}(\tilde A_{l+1}-\tilde A_{l})|< \varepsilon_l^{\frac23}$.
Then we can get, by (\ref{tilde_ZA_j+1}),
  \begin{equation}\label{sigma_m_j_tilde}
   \begin{array}{lll}
     |\tilde Z^{(M)}|_{(2\T)^d}\leq \varepsilon_j^{-\frac\sigma6}, & |\partial\tilde Z^{(M)}|_{(2\T)^d}\leq \varepsilon_j^{-\frac\sigma3},  & |\partial^2\tilde Z^{(M)}|_{(2\T)^d}\leq \varepsilon_j^{-\frac{\sigma}2},\\[1mm]
|\tilde A^{(M)}|\leq 5, & |\partial\tilde A^{(M)}|\leq N_j^{4\tau}, & |\partial^2\tilde A^{(M)}|\leq \varepsilon_j^{-\frac\sigma6}.
   \end{array}
  \end{equation}
Moreover, $|\xi^{(M)}|\leq |\xi_{j+1}|+2\varepsilon_{j}^{\frac14}\leq 2\varepsilon_{j}^{\sigma}$ on $\Gamma^{(M)}_{j+1}$, $0\leq j\leq J$.

As shown in the proof of Proposition \ref{propsana}, after the renormalization and the standard KAM regime, there may be one subinterval ${\cal I}\subset{\cal I}_{\la k_j^{(M)}\ra}$ on which $|{\rm tr}\tilde A^{(M)}|>2$. If $|{\rm tr}\tilde A^{(M)}|\leq2$ on ${\cal I}_{\la k_j^{(M)}\ra}$, then it can be seen as
 $|{\cal I}|=0$. We have $|\xi^{(M)}|\equiv0$ on ${\cal I}$, since it represents ``uniformly hyperbolic".
 We can refer to \cite{E92, HA} for more details.

On ${\cal I}_{\la k_j^{(M)}\ra}\setminus{\cal I}$, $\sin\xi^{(M)}$ does not vanish, and keeps the property $\partial\xi^{(M)}>\frac13$ as on $\Gamma^{(M)}_{0}$. Since $\partial\xi^{(M)}=-\frac{\partial {\rm tr}\tilde A^{(M)}}{2\sin\xi^{(M)}}$ for $\sin\xi^{(M)}\neq 0$, we can get (\ref{esti_plat}) by (\ref{sigma_m_j_tilde}).
\end{itemize}

Till now we have $|\tilde Z^{(M)}(\theta+\omega)^{-1}\, (A_0+F_0(\theta)) \,  \tilde Z^{(M)}(\theta)-\tilde A^{(M)}|_{(2\T)^d}\leq \varepsilon_{J+1}$.
 With $C_{\tilde A^{(M)}}$ the matrix of normalized eigenvectors of $\tilde A^{(M)}$, let  $\alpha^{(M)}:=\tilde\alpha^{(M)}+\sum_{j=0}^J\la k^{(M)}_j\ra$, and
$$ H^{(M)}(\theta):= C_{\tilde A^{(M)}} \left(\begin{array}{cc}
                          \exp\{-\frac{\rm i}{2}\sum_{j=0}^J\la k^{(M)}_j, \theta\ra\} & 0 \\[1mm]
                          0 & \exp\{\frac{\rm i}{2}\sum_{j=0}^J\la k^{(M)}_j, \theta\ra\}
                        \end{array}
\right) C_{\tilde A^{(M)}}^{-1},$$
and $A^{(M)}:= C_{\tilde A^{(M)}} \left(\begin{array}{cc}
                          e^{{\rm i}\alpha^{(M)}} & 0 \\
                          0 & e^{-{\rm i}\alpha^{(M)}}
                        \end{array}
\right) C_{\tilde A^{(M)}}^{-1}$, $Z^{(M)}:=\tilde Z^{(M)}\, H^{(M)}$.
Similar to (\ref{additional_transformation}), we can verify that
$$\tilde Z^{(M)}(\theta+\omega)\, \tilde A^{(M)} \,  \tilde Z^{(M)}(\theta)^{-1}=Z^{(M)}(\theta+\omega)\, A^{(M)} \, Z^{(M)}(\theta)^{-1}$$
So $e^{\pm{\rm i}\alpha^{(M)}}$ are the eigenvalues of $A^{(M)}$ and
$|\Re\alpha^{(M)}-\rho|= |\xi^{(M)}+\sum_{l=0}^{J}\la k_l^{(M)}\ra-\rho| \leq \varepsilon_J^{\frac14}$.
Since for $\sin\xi^{(M)}\neq 0$, $C_{\tilde A^{(M)}}$ is a normalization of $\left(\begin{array}{cc}
                      \tilde A^{(M)}_{12} & \tilde A^{(M)}_{12} \\
                      e^{{\rm i}\xi^{(M)}}-\tilde A^{(M)}_{11} &  e^{-{\rm i}\xi^{(M)}}-\tilde A^{(M)}_{11}
                    \end{array}\right)$, by a straightforward calculation, we can see, similar to (\ref{additional_procedure_H}) and (\ref{additional_procedure_B}),
\begin{equation}\label{H^M_A^M_H}
H^{(M)}(\theta):=\frac{\sin\frac{\sum_{j\geq0}\la k^{(M)}_j, \theta\ra}2}{\sin\xi^{(M)}}\left(\begin{array}{cc}
              \tilde A^{(M)}_{11} & \tilde A^{(M)}_{12} \\
              \tilde A^{(M)}_{21} & -\tilde A^{(M)}_{11}
            \end{array}\right) + \frac{\sin\left(\xi^{(M)}-\frac{\sum_{j\geq0}\la k^{(M)}_j, \theta\ra}2\right)}{\sin\xi^{(M)}}Id.,
\end{equation}
\begin{equation}\label{H^M_A^M_A}
A^{(M)}:=\frac{\sin\rho^{(M)}}{\sin\xi^{(M)}}\left(\begin{array}{cc}
              \tilde A^{(M)}_{11} & \tilde A^{(M)}_{12} \\
              \tilde A^{(M)}_{21} &  -\tilde A^{(M)}_{11}
            \end{array}\right) +\left(\begin{array}{cc}
                                         -\frac{\sin(\rho^{(M)}-\xi^{(M)})}{\sin\xi^{(M)}} & 0 \\
                                         0 &\frac{\sin(\rho^{(M)}+\xi^{(M)})}{\sin\xi^{(M)}}
                                       \end{array}
            \right).
\end{equation}

On $\Gamma^{(M)}_0$, $k^{(M)}_j=0$, $\forall \, 0\leq j\leq J$, so
$H^{(M)}= Id.$, $A^{(M)}=\tilde A^{(M)}$, $Z^{(M)} =\tilde Z^{(M)} $. Then (\ref{sigma_m_0}) for $\Gamma_0^{(M)}$ is proven in (\ref{sigma_m_0_tilde}).

On $\Gamma^{(M)}_{j+1}$, if $\sin\xi^{(M)}\neq 0$, then $\left|\partial^\nu \left(\frac{1}{\sin\xi^{(M)}}\right)\right|\leq N_j^{4\nu\tau}|\sin\xi^{(M)}|^{-(1+2\nu)}$, $\nu=1,2$, and
$$\begin{array}{llll}
  |\sin\xi^{(M)} H^{(M)}|_{(2\T)^d}, & |\sin\xi^{(M)} A^{(M)}| & \leq \;  5,&\\[1mm]
  |\partial^\nu(\sin\xi^{(M)} H^{(M)})|_{(2\T)^d}, & |\partial^\nu(\sin\xi^{(M)} A^{(M)})| & \leq \; 2 N_j^{4\nu\tau}|\sin\xi^{(M)}|^{1-2\nu},& \nu=1,2
\end{array} .$$
Combining all the estimates above, we get, for $\sin\xi^{(M)}\neq 0$,
$$
|\partial^\nu H^{(M)}|_{(2\T)^d},\;\ |\partial^{\nu} A^{(M)}|\leq N_j^{9\tau}|\sin\xi^{(M)}|^{-(1+2\nu)},\quad \nu=0,1,2.
$$
Then for $Z^{(M)}=H^{(M)} \tilde Z^{(M)}$, we have $|\partial^\nu Z^{(M)}|_{(2\T)^d}\leq \varepsilon_j^{-\frac\sigma5}|\sin\xi^{(M)}|^{-(1+2\nu)}$.
So (\ref{sigma_m_0}) is proven for $\Gamma_{j+1}^{(M)}$.

Since on $\Sigma_{j}$, $0\leq j \leq J+1$,
$$|\hat Z_{l+1}-Id.|_{(2\T)^d}< \varepsilon_l^{\frac12}, \;\ |\tilde A_{l+1}-\tilde A_{l}|< \varepsilon_l^{\frac23}, \quad l \geq J+1 ,$$
by the construction of $Z^{(M)}$, $A^{(M)}$, $Z$ and $B$, and noting that on $\Sigma_{j+1}\subset\Gamma^{(M)}_{j+1}$, $0\leq j\leq J$,
$|\sin\xi^{(M)}\, Z^{(M)}- \sin\xi\, Z|_{(2\T)^d}$ and $|\sin\xi^{(M)}\, A^{(M)}-\sin\xi\, B|$ are bounded by $|\tilde Z^{(M)}-\tilde Z|_{(2\T)^d}$ and
$|\tilde A^{(M)}-\tilde B|$ respectively, we can prove {\bf (S3)}.

\

\noindent{Proof of {\bf (S4)}:} From the construction of the intervals ${\cal I}_{\la k_j\ra}$, we can see $M\rho(\partial{\cal I}_{\la k_j\ra})\subset\pi\Z$ for $0\leq j\leq J-1$. By the definition of $\Gamma_{j}^{(M)}$ in (\ref{Gamma^M}), every connected component of $\Gamma_{J+1}^{(M)}$ is some ${\cal I}_{\la k_J\ra}$, so $\{E\in\partial\Gamma^{(M)}:M\rho(E)\notin\pi\Z\}\subset \partial\Gamma^{(M)}_{J+1}$.

Since every ${\cal I}_{\la k_J\ra}=(E_*, E_{**})$ is generated at the $(J+1)^{\rm th}-$step, it is contained in an interval in which $\tilde Z_j$ and $\tilde A_j$, $1\leq j\leq J$, are all ${\cal C}^2$,
and $k_{J}^{(M)}(E)=k_J$ on ${\cal I}_{\la k_J\ra}$, $k_{J}^{(M)}(E^-_*)=k_{J}^{(M)}(E^+_{**})=0$ since it is in the case ``non-resonance'' outside the interval ${\cal I}_{\la k_J\ra}$ at the $(J+1)^{\rm th}-$step.
Indeed, we can find $0 \leq j_* < J$, such that $E_*,\,  E_{**}\in \partial\Gamma^{(M)}_{j_*}$, and $k_{l}^{(M)}(E^-_*)=k_{l}^{(M)}(E^+_{**})=0$ for $j_*\leq l\leq J$.
For $\tilde Z_j=\prod_{l=j}^1\hat Z_{l}$, $1\leq j \leq J$, a similar computation as (\ref{tilde_ZA_j+1_0}) or (\ref{tilde_ZA_j+1}) shows that
$$\left\{\begin{array}{lll}
    \left|\partial(\tilde Z_{j}-Id.)\right|_{(2\T)^d}\leq \varepsilon_{0}^{\frac12}, & \left|\partial(\tilde A_{j}- A_0)\right|\leq  \varepsilon_{0}^{\frac23}, & j_*=0 \\[1mm]
    \left|\partial\tilde Z_{j}\right|_{(2\T)^d}\leq \varepsilon_{j_*-1}^{-\frac\sigma3}, & \left|\partial\tilde A_{j}\right|\leq N_{j_*-1}^{4\tau}, & j_*\geq 1
  \end{array} \right. .
$$
Moreover, outside the interval ${\cal I}_{\la k_J\ra}$,
$|\hat Z_{J+1}-Id.|_{(2\T)^d}$, $|\tilde A_{J+1} -\tilde A_{J}|\leq \varepsilon_J^{\frac12}$.
Recalling that $E_{**}-E_{*}\geq \varepsilon_J^{3\sigma(1+\sigma)}$,
we have
$\left|\left. \hat Z_{J+1}\right|_{E_*^-}^{E_{**}^+}\right|_{(2\T)^d} \leq \varepsilon_J^{\frac{5}{12}} \varepsilon_J^{3\sigma(1+\sigma)}\leq \varepsilon_J^{\frac{5}{12}} (E_{**}-E_{*})$ and
\begin{equation}\label{bd_tilde_AM}
\left\{\begin{array}{ll}
\left|\left.(\tilde A^{(M)}- A_0)\right|_{E_*^-}^{E_{**}^+}\right|\leq 3\varepsilon_{0}^{\frac12}(E_{**}-E_{*}), & j_*=0 \\ [1mm]
           \left|\left. \tilde A^{(M)} \right|_{E_*^-}^{E_{**}^+}\right| \leq 3N_{j_*-1}^{4\tau} (E_{**}-E_{*}),& j_*\geq 1
         \end{array}
    \right. .
\end{equation}
For $\left.\tilde Z^{(M)}\right|_{E_*^-}^{E_{**}^+}=\left.\hat Z_{J+1}\right|_{E_*^-}^{E_{**}^+}\cdot \tilde Z_J(E_{**})+\hat Z_{J+1}(E_{*}^-)\cdot \left.\tilde Z_J\right|_{E_*^-}^{E_{**}^+}$, we have
\begin{equation}\label{bd_tilde_ZM}
\left|\left.\tilde Z^{(M)}\right|_{E_*^-}^{E_{**}^+}\right|_{(2\T)^d}\leq \left\{\begin{array}{ll}
           \displaystyle \frac12\varepsilon_0^{\frac13} (E_{**}-E_{*}), & j_*=0 \\[2mm]
           3\varepsilon_{j_*-1}^{-\frac{\sigma}{3}} (E_{**}-E_{*}),& j_*\geq 1
         \end{array}
    \right. .
\end{equation}

The estimate (\ref{edge_point}) for $j_*=0$ follows directly from (\ref{bd_tilde_AM}) and (\ref{bd_tilde_ZM}), by noting that $\left. Z^{(M)}\right|^{E^+_{**}}_{E^-_*}=\left.\tilde Z^{(M)}\right|^{E^+_{**}}_{E^-_*}$ and $\left. A^{(M)} \right|^{E^+_{**}}_{E^-_*}=\left.\tilde A^{(M)}\right|^{E^+_{**}}_{E^-_*}$.
For $j_*\geq 1$, by noting that $\partial\xi_J=\partial\rho_J=-\frac{\partial{\rm tr}\tilde A_J}{2\sin\xi_J}$ and $|\sin\xi_J| \leq 2\varepsilon_{j_*-1}^{\sigma}$,
$$\left|\left.\sin^3\xi^{(M)}\, \sin\rho^{(M)} \right|_{E_*^-}^{E_{**}^+}\right|
\leq \left|\left.\sin^3\xi_J\, \sin\rho_J \right|_{E_*}^{E_{**}}\right|+2 \varepsilon_J^{\frac14}\leq N_{j_*-1}^{5\tau}\varepsilon_{j_*-1}^{\sigma} (E_{**}-E_{*}).$$
Similarly, $\left|\left.\sin^3\xi^{(M)}\, \sin\left(\xi^{(M)}-\frac{\sum_{j\geq0}\la k^{(M)}_j, \theta\ra}2\right) \right|_{E_*^-}^{E_{**}^+}\right|$, $\left|\left.\sin^3\xi^{(M)} \sin\left(\rho^{(M)}\pm\xi^{(M)}\right) \right|_{E_*^-}^{E_{**}^+}\right|$, and $\left|\left.\sin^3\xi^{(M)} \right|_{E_*^-}^{E_{**}^+}\right|$ are all bounded by $N_{j_*-1}^{5\tau}\varepsilon_{j_*-1}^{\sigma}(E_{**}-E_{*})$.
In view of (\ref{H^M_A^M_H}) and (\ref{H^M_A^M_A}),
  \begin{eqnarray*}
\left|\left.\sin^4\xi^{(M)}\, A^{(M)}\right|^{E^+_{**}}_{E^-_*}\right|
&\leq& \left|\left.\sin^3\xi^{(M)} \sin \rho^{(M)}\tilde A^{(M)}\right|^{E^+_{**}}_{E^-_*}\right|+ \left|\left.\sin^3\xi^{(M)} \sin\left(\rho^{(M)}\pm\xi^{(M)}\right)\right|^{E^+_{**}}_{E^-_*}\right|\\
&\leq&\left(4N_{j_*-1}^{5\tau}\varepsilon^\sigma_{j_*-1}+ 2\varepsilon^\sigma_{j_*-1}\cdot 3N_{j_*-1}^{4\tau}+N_{j_*-1}^{5\tau}\varepsilon_{j_*-1}^{\sigma}\right) (E_{**}-E_{*})\\
&\leq& \varepsilon_{j_*-1}^{\frac{3\sigma}{4}}  (E_{**}-E_{*}),
  \end{eqnarray*}
  and similarly $\left|\left.\sin^4\xi^{(M)}\, H^{(M)}\right|^{E^+_{**}}_{E^-_*}\right|_{(2\T)^d}\leq \varepsilon_{j_*-1}^{\frac{3\sigma}{4}} (E_{**}-E_{*})$.
  Finally, for $Z^{(M)}=H^{(M)}\tilde Z^{(M)}$
$$\left|\left.\sin^4\xi^{(M)}\, Z^{(M)}\right|^{E^+_{**}}_{E^-_*}\right|_{(2\T)^d}=\left|\left.\sin^4\xi^{(M)}\, H^{(M)}\tilde Z^{(M)}\right|^{E^+_{**}}_{E^-_*}\right|_{(2\T)^d}\leq \frac12 \varepsilon_{j_*-1}^{\frac{2\sigma}{3}} (E_{**}-E_{*}).$$
\qed

\subsection{Proof of Proposition \ref{propsana}(3)}\label{proofP13}
\noindent

Recall $\breve{Z}$ and $\breve B$ given in the proof of Proposition \ref{propsana}(1) and (2), and the expressions of $H$ and $B$ given in (\ref{additional_procedure_H}) and (\ref{additional_procedure_B}).
We also define
\begin{eqnarray*}
 \hat H(\theta)&:=&\frac{\sin\frac{\sum_{j\geq0}\la k_j, \theta\ra}2}{\sin\xi}\left(\begin{array}{cc}
              \breve B_{11} & \breve B_{12} \\
              \breve B_{21} & -\breve B_{11}
            \end{array}\right)  + \frac{\sin\frac{\sum_{j\geq0}\la k_j, \theta\ra}2\cos\xi\cdot{\rm tr}\breve B}{2\sin^3\xi}\left(\begin{array}{cc}
              \tilde B_{11} & \tilde B_{12} \\
              \tilde B_{21} & -\tilde B_{11}
            \end{array}\right)\\[1mm]
 &  &  + \frac{\sin\left(\xi-\frac{\sum_{j\geq0}\la k_j, \theta\ra}2\right)\cos\xi\cdot{\rm tr}\breve B}{2\sin^3\xi}Id.  - \frac{\cos\left(\xi-\frac{\sum_{j\geq0}\la k_j, \theta\ra}2\right)\cdot{\rm tr}\breve B}{2\sin^2\xi}Id. ,  \\[1mm]
 \hat B&:=&
 \frac{\sin\rho}{\sin\xi}\left(\begin{array}{cc}
              \breve B_{11} & \breve B_{12} \\
              \breve B_{21} &  -\breve B_{11}
            \end{array}\right) +\left(\frac{\sin\rho\cos\xi\cdot{\rm tr}\breve B }{2\sin^3\xi}-\frac{\cos\rho\cdot{\rm tr}\breve B }{2\sin^2\xi}\right) \left(\begin{array}{cc}
              \tilde B_{11} & \tilde B_{12} \\
              \tilde B_{21} & -\tilde B_{11}
            \end{array}\right) \\[1mm]
 & &     +\frac{\cos\xi\cdot{\rm tr}\breve B }{2\sin^3\xi}\left(\begin{array}{cc}
-\sin(\rho-\xi) & 0 \\
0 & \sin(\rho+\xi)
\end{array}\right)  -\frac{{\rm tr}\breve B}{\sin^2\xi}\left(\begin{array}{cc}
0 & 0 \\
 0 & \cos(\rho+\xi)
\end{array}\right).
\end{eqnarray*}
$\hat H$ and $\hat B$ can be (formally) seen as the derivative of $H$ and $B$ respectively. In particular, $\hat H=0$ and $\hat B=\breve B$ on $\Sigma_0$.

%
We are going to show that
\begin{itemize}
\item on $\Sigma_0$, $Z$ and $B$ are ${\cal C}^1_W$ with the first order derivatives $\bar Z:=\breve{Z}$ and $\bar B:=\breve B$ respectively;
\item on $\Sigma_{j+1}$, $\sin^{s+2}\xi\cdot Z$ and $\sin^{s+2}\xi\cdot B$, $s\geq2$, are ${\cal C}^1_W$ with the first order derivatives
\begin{eqnarray*}
\bar Z&:=&\sin^{s+2}\xi\, (\tilde Z\cdot \hat{H}+\breve Z\cdot H)-\frac{(s+2){\rm tr}\breve B}{2}\sin^{s}\xi\cos\xi\cdot Z, \\
\bar B&:=&\sin^{s+2}\xi\cdot \hat B-\frac{(s+2){\rm tr}\breve B}{2}\sin^{s}\xi\cos\xi \, \cdot B.
\end{eqnarray*}
\end{itemize}
By the estimates in (\ref{tilde_ZB}) and the above expressions of $\hat H$ and $\hat B$, we have
$$
\left\{\begin{array}{ll}
|\bar Z|_{\Sigma_0,\, (2\T)^d}, & |\bar B-\partial A_0|_{\Sigma_0}\leq 2\varepsilon_0^{\frac12}  \\[1mm]
 |\bar Z|_{\Sigma_{j+1},\, (2\T)^d}, & |\bar B|_{\Sigma_{j+1}}\leq \frac14 \varepsilon_j^{\frac{2\sigma}{3}}
\end{array}\right. ,
$$
since $0< |\xi| \leq  2 \varepsilon_j^{\sigma}$ on $\Sigma_{j+1}$.
In view of Definition \ref{definition_whitney}, to finish the proof of (\ref{limit_state_whitney}), it is sufficient to show that, for any $E_1$, $E_2\in\Sigma_j$ with $E_1<E_2$, and for $s\geq2$,
 \begin{equation}\label{edge_point_whitney}
\left\{\begin{array}{llll}
\left|\left. (Z-Id.)\right|_{(E_1, E_2)}\right|_{(2\T)^d},   & \left|\left. (B-A_0)\right|_{(E_1, E_2)}\right|   &  \leq \frac{1}{2}\varepsilon_0^{\frac13}(E_{2}-E_1), & j=0 \\[3mm]
\left|\left.\sin^{s+2}\xi\cdot Z\right|_{(E_1, E_2)} \right|_{(2\T)^d},   & \left|\left.\sin^{s+2}\xi\cdot B\right|_{(E_1, E_2)}\right|  &  \leq  \frac{1}{2}\varepsilon_{j-1}^{\frac{2\sigma}{3}}(E_{2}-E_1), & j\geq1
\end{array}\right. .
\end{equation}
Since, in view of (\ref{ZB}), (\ref{edge_point_whitney}) is evident if $E_2-E_1> \varepsilon_{j}^\sigma$, we assume that $E_2-E_1 \leq \varepsilon_{j}^\sigma$,

There is some $J\geq j$ such that $\varepsilon_{J+1}^{\sigma}\leq E_2-E_1\leq \varepsilon_J^{\sigma}$. So we can choose $M\in\Z\setminus\{0\}$ with $J(M)=J$ such that $\Sigma_j\subset \Gamma_j^{(M)}$, and let $\tilde Z^{(M)}=\tilde Z_{J+1}$, $\tilde A^{(M)}=\tilde A_{J+1}$ be constructed as in the proof of Proposition \ref{propsana1}.
Then we have
\begin{equation}\label{error_whitney}
|\tilde Z-\tilde Z^{(M)}|_{\Sigma_j, (2\T)^d},\; |\tilde B-\tilde A^{(M)}|_{\Sigma_j}, \; |\xi-\xi^{(M)}|_{\Sigma_j} \leq \varepsilon_{J}^{\frac13-\sigma(1+\sigma)} (E_2-E_1).
\end{equation}
If $(E_1, E_2)\subset\Gamma_j^{(M)}$, then $\tilde Z^{(M)}$, $\tilde A^{(M)}$ are ${\cal C}^2$ on $(E_1, E_2)$. So, by (\ref{sigma_m_0_tilde}) and (\ref{sigma_m_j_tilde}),
$$\left\{\begin{array}{llll}
\left| \left.(\tilde Z^{(M)}-Id.) \right|_{(E_1, E_2)} \right|_{(2\T)^d},&\left|\left.(\tilde A^{(M)}-A_0) \right|_{(E_1, E_2)} \right|&\displaystyle \leq 2\varepsilon_0^{\frac12}(E_2-E_1),&  j=0\\[3mm]
\left| \left.\tilde Z^{(M)} \right|_{(E_1, E_2)} \right|_{(2\T)^d},&\left|\left.\tilde A^{(M)} \right|_{(E_1, E_2)} \right|&\leq \varepsilon_{j-1}^{-\frac{\sigma}6}(E_2-E_1), &  j\geq 1
\end{array}\right. ,$$
and, by (\ref{esti_plat}) , $\left|\left.\sin^{s+1} \xi^{(M)}\right|_{(E_1, E_2)} \right|\leq \varepsilon_{j-1}^{\frac{5\sigma}6}(E_2-E_1)$
for any given $s\geq2$. So, by (\ref{H^M_A^M_H}) and (\ref{H^M_A^M_A}), we have
\begin{equation}\label{edge_point_whitney_M}
\left\{\begin{array}{llll}
\left|\left.(Z^{(M)}-Id.)\right|_{(E_1, E_2)}\right|_{(2\T)^d},   & \left|\left.(A^{(M)}-A_0)\right|_{(E_1, E_2)}\right|   &  \displaystyle \leq \frac12\varepsilon_0^{\frac13}(E_{2}-E_1), & j=0 \\[3mm]
\left|\left.\sin^{s+2}\xi^{(M)}\, Z^{(M)}\right|_{(E_1, E_2)} \right|_{(2\T)^d},   & \left|\left.\sin^{s+2}\xi^{(M)}\, A^{(M)}\right|_{(E_1, E_2)}\right|  &  \displaystyle \leq \frac12 \varepsilon_{j-1}^{\frac{2\sigma}{3}}(E_{2}-E_1), & j\geq1
\end{array}\right.  .
\end{equation}
If there is a subset $S\subset(E_1, E_2)$ but $S\cap\Gamma_j^{(M)}=\emptyset$, then it must be the union of
 connected components of $\Gamma_{J+1}^{(M)}$ since $\varepsilon_{J+1}^{\sigma}\leq E_2-E_1\leq \varepsilon_J^{\sigma}$. For any connected component $(E_*, E_{**})$ of $\Gamma_{J+1}^{(M)}$,
 (\ref{edge_point}) implies the same estimates as (\ref{edge_point_whitney_M}) between $E_*^-$ and $E_{**}^+$. So, combining with (\ref{error_whitney}), we can get (\ref{edge_point_whitney}). \qed

\

\noindent {\bf Acknowledgements.} The author would like to thank H. Eliasson for many fruitful discussions about this work, and
J.-C. Yoccoz for supporting this work during the period of the post-doc contract of ANR(L'Agence nationale de la recherche).
He would also thank D. Damanik and J. You for pointing out some references and giving important advises.
The author would also thank the anonymous referees for valuable comments and suggestions on the manuscript.

\end{document}